\documentclass[prd,twocolumn,eqsecnum,floatfix]{revtex4}
\usepackage{bm}
\usepackage{amssymb}
\usepackage{graphicx}
\begin{document}
\title{Low multipole contributions to the gravitational self-force}
\author{Steven Detweiler}
\affiliation{Department of Physics, PO Box 118440, University of Florida,
Gainesville, FL 32611-8440}
\author{Eric Poisson}
\affiliation{Department of Physics, University of Guelph, Guelph,
Ontario, Canada N1G 2W1}
\date{December 1, 2003} 
\begin{abstract}

We calculate the unregularized monopole and dipole contributions to
the self-force acting on a particle of small mass in a circular orbit
around a Schwarzschild black hole. From a self-force point of view,
these non-radiating modes are as important as the radiating modes with
$l\ge2$. In fact, we demonstrate how the dipole self-force contributes
to the dynamics even at the Newtonian level.

The self-acceleration of a particle is an inherently gauge-dependent
concept, but the Lorenz gauge is often preferred because of its
hyperbolic wave operator. Our results are in the Lorenz gauge and are
also obtained in closed form, except for the even-parity dipole case
where we formulate and implement a numerical approach.

\end{abstract}
\pacs{04.25.-g,  04.40.-b, 04.20.-q, 04.70.Bw, 97.60Lf}
\maketitle

\section{Introduction}

The capture of solar-mass compact objects by massive black holes
residing in galactic centers has been identified as one of the most
promising sources of gravitational waves for the Laser Interferometer
Space Antenna \cite{LISA}. The need for accurate templates for signal
detection and source identification is currently motivating an
intense effort from many workers to determine the motion of a
relativistic two-body system in the small mass-ratio limit, without
relying on slow-motion or weak-field approximations; for a review,
see Ref.~\cite{poisson:03}. The work presented in this paper is part
of this larger effort.

\subsection*{Gravitational self-force and the MiSaTaQuWa
equations of motion}

Consider a small body of mass $m$ in orbit around a much larger black
hole of mass $M$. In the test-mass limit ($m \to 0$) the motion of
the small body is known to follow a geodesic in the spacetime
geometry of the large black hole \cite{manasse:63, kates:80,
thorne-hartle:85, death:96, ehlers-geroch:03}. But as the mass of the
smaller object is allowed to increase, deviations from geodesic
motion become noticeable; these are associated with important
physical effects such as radiation reaction and finite-mass
(conservative) corrections to the orbital motion. In a sense, the
motion is now geodesic in the perturbed spacetime that contains both
the black hole and the orbiting body. If $g_{\alpha\beta}$ denotes
the unperturbed metric of the central black hole, and if
$h_{\alpha\beta}$ denotes the perturbation produced by the orbiting
body, then the motion is formally geodesic in the perturbed metric
$g_{\alpha\beta} + h_{\alpha\beta}$. When viewed from the background
spacetime, the motion appears to be accelerated, and the agent that
produces the acceleration is a gravitational self-force acting on the
particle.

To turn these considerations into concrete equations of motion, it is
desirable to formulate an approximation in which the details of the
small body's internal structure have a negligible influence on the
body's orbital motion. In this approximation the body is modeled as a
point particle possessing mass but no higher multipole moments, and
its motion is fully described in terms of a world line $\gamma$. But
to formulate equations of motion for this world line becomes
problematic, as the field $h_{\alpha\beta}$ produced by a point
particle necessarily diverges at the position of the particle. This
means that an affine connection cannot be defined on the world line,
and that the statement ``the particle follows a geodesic of the
perturbed spacetime'' does not make immediate sense.

The task of regularizing $h_{\alpha\beta}$ near the world line and
formulating meaningful equations of motion for the point particle was
undertaken by Mino, Sasaki, and Tanaka \cite{mino-etal:97}, and also
by Quinn and Wald \cite{quinn-wald:97}. An interesting reformulation
of this work was recently given by Detweiler and Whiting
\cite{detweiler-whiting:03}, who showed that the perturbation can be
uniquely decomposed into a symmetric-singular field $h^{\rm
S}_{\alpha\beta}$, and a regular-radiative field $h^{\rm
R}_{\alpha\beta}$; the full (retarded) perturbation is then
$h_{\alpha\beta} = h^{\rm S}_{\alpha\beta} + h^{\rm
R}_{\alpha\beta}$. Detweiler and Whiting were able to establish that
while $h^{\rm S}_{\alpha\beta}$ reproduces the singularity
structure of the metric perturbation, it exerts no force on the point
particle; the gravitational self-force is then produced entirely by
$h^{\rm R}_{\alpha\beta}$, which is a homogeneous, regular, smooth
field in a neighbourhood of the world line.

The MiSaTaQuWa equations of motion \cite{mino-etal:97, quinn-wald:97}, in the
Detweiler-Whiting formulation \cite{detweiler-whiting:03}, take the following
form. Let $z^\mu(\tau)$ be parametric relations that describe the particle's
world line $\gamma$, with $\tau$ denoting proper time in the background
spacetime of the central black hole. Let $u^\mu = dz^\mu/d\tau$ be the
particle's four-velocity, normalized with respect to the unperturbed metric:
$g_{\mu\nu} u^\mu u^\nu = -1$. Let $D/d\tau$ denote covariant differentiation
along the world line, defined with respect to a connection compatible with
$g_{\alpha\beta}$. Then the particle's equations of motion are
\begin{eqnarray}
\frac{D u^\mu}{d\tau} &=& a^\mu\bigl[ h^{\rm R} \bigr]
\nonumber \\
&\equiv& -\frac{1}{2} \bigl( g^{\mu\nu} + u^\mu
u^\nu \bigr) \bigl( 2 h^{\rm R}_{\nu\lambda;\rho}
- h^{\rm R}_{\lambda\rho;\nu} \bigr) u^\lambda u^\rho,
\label{1.1}
\end{eqnarray}
where a semicolon indicates covariant differentiation with respect to
the background connection. The right-hand side of Eq.~(\ref{1.1}) is
the gravitational self-acceleration of the point particle; multiplying
by $m$ would give the gravitational self-force. Equation (\ref{1.1})
is equivalent to the statement that the particle moves on a
geodesic in a spacetime with metric $g_{\alpha\beta}
+ h^{\rm R}_{\alpha\beta}$, but the description of the world line
refers to the background spacetime. The right-hand side of
Eq.~(\ref{1.1}) is of order $m$, and the gravitational
self-acceleration is therefore $O(m)$; it vanishes in the test-mass
limit and the motion becomes geodesic (in the background spacetime).

The decomposition of $h_{\alpha\beta}$ into singular ``S'' and
radiative ``R'' fields relies on a specific choice of gauge for the
metric perturbation, which must satisfy the Lorenz gauge condition
\begin{equation}
\nabla_\beta \Bigl(
h^{\alpha\beta} - \frac{1}{2} g^{\alpha\beta} g^{\gamma\delta}
h_{\gamma\delta} \Bigr) = 0.
\label{1.2}
\end{equation}
This choice ensures that $h_{\alpha\beta}$ satisfies a (hyperbolic)
wave equation and that the correct, retarded solution can be
identified. The singularity structure of the perturbation near the
world line can then be determined by a local analysis (see, for
example, Ref.~\cite{poisson:03}), and $h^{\rm S}_{\alpha\beta}$ is
constructed without ambiguity so that it exerts no force on the
particle. The regular field $h^{\rm R}_{\alpha\beta}$ is then the
difference between the retarded solution and the locally-constructed
singular field; this satisfies a homogeneous version of the wave
equation satisfied by the full metric perturbation, and the metric
$g_{\alpha\beta} + h^{\rm R}_{\alpha\beta}$ is a solution to the
linearized Einstein field equations in vacuum.

The Lorenz gauge therefore presents itself as a preferred gauge for
this problem, and it has been shown that in the first post-Newtonian
approximation, Eq.~(\ref{1.1}) agrees with the standard
Einstein-Infeld-Hoffmann equations of motion in a common
domain of validity \cite{pfenning-poisson:02}. But it is important to
note that the equations of motion of Eq.~(\ref{1.1}) are not gauge
invariant \cite{barack-ori:01}: different gauge choices will lead to
different results.

\subsection*{Self-acceleration by mode sums}

A concrete evaluation of Eq.~(\ref{1.1}) is challenging and involves
a large number of steps; for this discussion we consider the case of a
particle orbiting a Kerr black hole.

The first sequence of steps are concerned with the computation of the
metric perturbation $h_{\alpha\beta}$ produced by a point
particle moving on a specified geodesic of the Kerr spacetime. A
method for doing this was elaborated by Lousto and Whiting
\cite{lousto-whiting:02} and Ori \cite{ori:03}, building on the
pioneering work of Teukolsky \cite{teukolsky:73}, Chrzanowski
\cite{chrzanowski:75}, and Wald \cite{wald:78}. The procedure consists
of (i) solving the Teukolsky equation for one of the Newman-Penrose
quantities $\psi_0$ and $\psi_4$ (complex components of the Weyl
tensor) produced by the point particle; (ii) obtaining from $\psi_0$
or $\psi_4$ a related (Hertz) potential $\Psi$ by integrating an
ordinary differential equation; (iii) applying to $\Psi$ a number of
differential operators to obtain the metric perturbation in a
radiation gauge that differs from the Lorenz gauge; and (iv)
performing a gauge transformation from the radiation gauge to
the Lorenz gauge.

It is well known that the Teukolsky equation can be separated when
$\psi_0$ or $\psi_4$ is expressed as a multipole expansion, summing
over modes with (spheroidal-harmonic) indices $l$ and $m$. In fact,
the procedure outlined above relies heavily on this mode
decomposition, and the metric perturbation returned at the end of the
procedure is also expressed as a sum over modes
$h^l_{\alpha\beta}$. (For each $l$, $m$ ranges from $-l$ to $l$, and
summation of $m$ over this range is henceforth understood.) From
these, mode contributions to the self-acceleration can be computed:
$a^\mu[h_l]$ is obtained from Eq.~(\ref{1.1}) by substituting
$h^l_{\alpha\beta}$ in place of $h^{\rm R}_{\alpha\beta}$. These mode
contributions do not diverge on the world line, but $a^\mu[h_l]$ is
discontinuous at the radial position of the orbit. The sum over modes,
on the other hand, does not converge, because the ``bare''
acceleration (constructed from the retarded field $h_{\alpha\beta}$)
is formally infinite.

The next sequence of steps is concerned with the regularization of
each $a^\mu[h_l]$ by removing the contribution from $h^{\rm
S}_{\alpha\beta}$ \cite{barack-etal:02, barack-ori:02,
barack-ori:03a, barack-ori:03b, mino-etal:03, detweiler-etal:03}. The
singular field can be constructed locally in a neighbourhood of the
particle, and then decomposed into modes of multipole order $l$. This
gives rise to modes $a^\mu[h^{\rm S}_l]$ for the singular part of the
self-acceleration; these are also finite and discontinuous, and their
sum over $l$ also diverges. But the true modes $a^\mu[h^{\rm R}_l] =
a^\mu[h_l] - a^\mu[h^{\rm S}_l]$ of the self-acceleration are
continuous at the radial position of the orbit, and their sum does
converge to the particle's acceleration. (It should be noted that
obtaining a mode decomposition of the singular field involves
providing an extension of $h^{\rm S}_{\alpha\beta}$ on a sphere of
constant radial coordinate, and then integrating over the angular
coordinates. The arbitrariness of the extension introduces
ambiguities in each $a^\mu[h^{\rm S}_l]$, but the ambiguity
disappears after summing over $l$.)

The gravitational self-acceleration is thus obtained by first
computing $a^\mu[h_l]$ from the metric perturbation derived
from $\psi_0$ or $\psi_4$, then computing the counterterms
$a^\mu[h^{\rm S}_l]$ by mode-decomposing the singular field, and
finally summing over all $a^\mu[h^{\rm R}_l] = a^\mu[h_l]
- a^\mu[h^{\rm S}_l]$. This procedure is lengthy and involved, and
thus far it has not been brought to completion, except for the special
case of a particle falling radially toward a nonrotating black hole
\cite{barack-lousto:02}. In this regard it should be noted that
replacing the central Kerr black hole by a Schwarzschild black hole
simplifies the task considerably. In particular, because there exists
a practical and well-developed formalism to describe the metric
perturbations of a Schwarzschild spacetime \cite{regge-wheeler:57,
zerilli:70, moncrief:74, sago-etal:03, jhingan-tanaka:03}, there is no
necessity to rely on the Teukolsky formalism and the complicated
reconstruction of the metric variables.

\subsection*{Low multipoles --- this work}

The procedure described above is not complete. The reason is that the
metric perturbations $h^l_{\alpha\beta}$ that can be recovered from
$\psi_0$ or $\psi_4$ do not by themselves sum up to the complete
gravitational perturbation produced by the moving particle. Missing
are the perturbations derived from the other Newman-Penrose
quantities: $\psi_1$, $\psi_2$, and $\psi_3$. While $\psi_1$ and
$\psi_3$ can always be set to zero by an appropriate choice of null
tetrad, $\psi_2$ contains such important physical information as the
shifts in mass and angular-momentum parameters produced by the
particle \cite{wald:73}. Because the mode decompositions of $\psi_0$
and $\psi_4$ start at $l=2$, we might colloquially say that what is
missing from the above procedure are the ``$l=0$ and $l=1$''
components of the metric perturbations. It is not currently known how
the procedure can be completed so as to incorporate {\it all
components} of the metric perturbations.

In this paper we consider the contribution of these low multipoles
($l=0$ and $l=1$) to the gravitational self-acceleration. To make
progress we shall take the central black hole to be nonrotating, and
the metric of the background spacetime to be a Schwarzschild
solution. This simplification allows us to use the robust formalism
of gravitational perturbations of the Schwarzschild spacetime
\cite{regge-wheeler:57, zerilli:70, moncrief:74, sago-etal:03,
jhingan-tanaka:03}, and more importantly, to define precisely what is
meant by the ``$l=0$ and $l=1$'' modes of the perturbation field. In
this context the associations between the $l=0$ mode and a shift of
mass parameter, the odd-parity $l=1$ mode and a shift of
angular-momentum parameter, and the reduction of the even-parity
$l=1$ mode to a gauge transformation, were first established by
Zerilli \cite{zerilli:70}. These associations are central to our
discussion, and we believe that the results derived here will have a
direct counterpart in the case of a Kerr black hole: The missing
metric perturbations of the Kerr spacetime will be equivalent to our
$l=0$ and $l=1$ perturbation modes in the limit where the black-hole
angular momentum goes to zero.

To keep our discussion concrete and the mathematical complexities to
a minimum, we calculate the $l=0$ and $l=1$ perturbation modes for
the specific case of a particle moving on a circular orbit of radius
$R$ and angular velocity $\Omega = \sqrt{M/R^3}$. While finding
solutions to the relevant perturbation equations can be a simple task
when adopting a simple choice of gauge (as we shall see), we insist
here, for reasons that were listed before, that the $l=0$ and $l=1$
perturbation modes should be calculated in the Lorenz gauge. This
complicates the structure of the perturbation equations, and finding
solutions is more challenging. We nevertheless are able to find exact
analytical solutions for the cases $l=0$ and $l=1$ (odd parity). For
even-parity $l=1$, however, we have to rely on numerical methods for
exact results, and a post-Newtonian approximation for analytical
results.

In the remaining sections of the paper we calculate the contributions
\[
a^\mu[h_{l=0}], \quad
a^\mu[h_{l=1;{\rm odd\ parity}}], \quad
a^\mu[h_{l=1;{\rm even\ parity}}]
\]
to the ``bare'' self-acceleration of a particle moving on a circular
orbit around a Schwarzschild black hole. Our expressions are finite
but discontinuous at the radial position of the particle: the answers
obtained when approaching $r=R$ from the interior ($r<R$), and those
obtained from exterior ($r>R$), do not match. We find in all cases
that the contribution to the bare acceleration is purely radial:
$a^\mu[h_l] = a^r[h_l] \delta^\mu_r$ for all three modes considered
here. Moreover, in all cases the self-acceleration is conservative and
does not contribute to the radiation reaction.

To keep the notation simple we shall set
\begin{eqnarray}
a[l=0] &\equiv& a^r[h_{l=0}], \nonumber \\
a[l=1; \mbox{odd}] &\equiv& a^r[h_{l=1;{\rm odd\ parity}}],
\label{1.3} \\
a[l=1; \mbox{even}] &\equiv& a^r[h_{l=1;{\rm even\ parity}}].
\nonumber
\end{eqnarray}
We display our results for $a[l](R)$ in two figures. In
Fig.~1 we show the results as calculated from the orbit's interior
($r \to R^-$) and in Fig.~2 we show our results as calculated from
the orbit's exterior ($r \to R^+$). These results do not have an
immediate physical meaning. To produce meaning they must be included
with higher-multipole contributions in a sum over all modes. Because
there exists no procedure to uniquely remove the ``S part'' of the
$l=0$ and $l=1$ perturbation modes (as was mentioned previously), we
are not able here to produce expressions for the low-multipole
contributions to the regularized self-acceleration.

\begin{figure}
\includegraphics[angle=-90,scale=0.33]{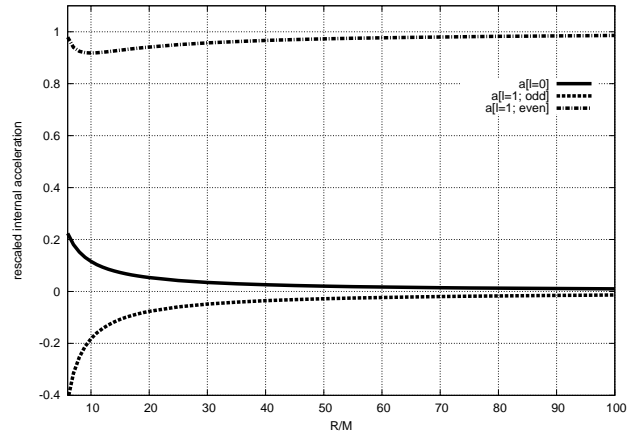}
\caption{Internal values of $a[l](R)$, rescaled by a common factor of
$3 m/R^2$. For $R \gg M$ we have the following asymptotic behaviors:
$a_<[l=0] \sim 3(m/R^2)(M/R)$, $a_<[l=1;\mbox{odd}] \sim
-4(m/R^2)(M/R)$, and $a_<[l=1;\mbox{even}] \sim 3(m/R^2)$. An exact
expression for $a_<[l=0]$ appears in Eq.~(\ref{3.15}) below. An exact
expression for $a_<[l=1;\mbox{odd}]$ appears in Eq.~(\ref{4.2}). The
values for $a_<[l=1;\mbox{even}]$ are obtained from Eq.~(\ref{5.48})
and the results listed in Table I.}
\end{figure}

\begin{figure}
\includegraphics[angle=-90,scale=0.33]{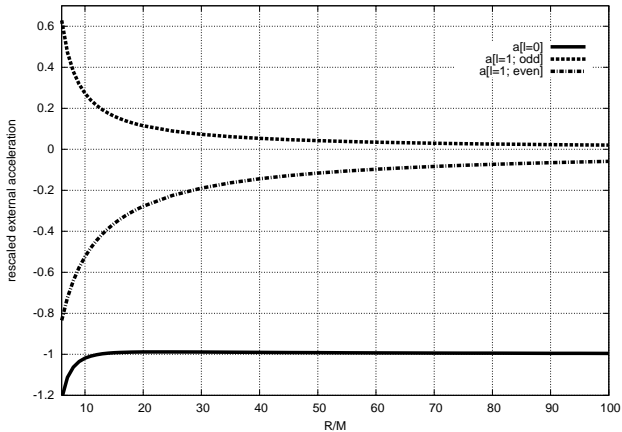}
\caption{External values of $a[l](R)$, rescaled by a common factor of
$m/R^2$. For $R \gg M$ we have the following asymptotic behaviors:
$a_>[l=0] \sim -m/R^2$, $a_<[l=1;\mbox{odd}] \sim 2(m/R^2)(M/R)$, and
$a_<[l=1;\mbox{even}] \sim -3\beta(m/R^2)(M/R)$, where $\beta \simeq
2$ is numerically estimated at the end of Sec.~V. An exact expression
for $a_>[l=0]$ appears in Eq.~(\ref{3.16}) below. An exact expression
for $a_>[l=1;\mbox{odd}]$ appears in Eq.~(\ref{4.3}). The values for
$a_>[l=1;\mbox{even}]$ are obtained from Eq.~(\ref{5.48}) and the
results listed in Table I.}
\end{figure}

Using purely analytical methods, Nakano, Sago and Sasaki
\cite{nakano-etal:03} calculated the self-acceleration to first
post-Newtonian order for circular orbits of the Schwarzschild
geometry. For the even and odd parity $l=1$ modes, their results for
the contribution to the ``bare'' self-acceleration agree with ours at
the 1PN level, as expected. It appears that an extension of their
methods to higher post-Newtonian orders might be substantially
complicated by the difficulty caused by the even-parity $l=1$
perturbations, the case for which we have to rely on numerical
methods. For the $l=0$ mode our results for the ``bare''
self-accelerations disagree with theirs at 1PN. We believe that the
discrepancy is caused by the implementation of boundary conditions at
the event horizon, and we discuss this matter in some detail in the
Appendix.

\subsection*{Organization of this paper}

In Sec.~II we set the stage with a discussion of the gravitational
self-force in Newtonian theory. This simple analogue to the
relativistic problem sheds considerable light on the meaning of the
self-acceleration and its decomposition into singular ``S'' and
regular ``R'' fields. We show in particular that in Newtonian theory,
the $l=1$ contribution to the self-acceleration is responsible for an
important finite-mass correction to the particle's angular velocity.
We take this as a clear suggestion that in the relativistic problem,
the low multipole contributions to the gravitational
self-acceleration produce important physical effects.

In Sec.~III we compute, in the Lorenz gauge, the $l=0$ gravitational
perturbations produced by a particle in a circular orbit around a
Schwarzschild black hole. These perturbations are associated with the
change of mass parameter that occurs at $r=R$. We then calculate
$a[l=0]$, the corresponding contribution to the self-acceleration.

In Sec.~IV we do the same for the $l=1$, odd-parity
perturbations. These are associated with the change of
angular-momentum parameter that occurs across the orbit, and they give
rise to the contribution $a[l=1; \mbox{odd}]$ to the
self-acceleration.

In Sec.~V we consider the $l=1$, even-parity gravitational
perturbations, which are associated with the motion of the central
black hole around the system's center of mass. This calculation
is considerably more involved than the others, because here the source
of the perturbations is time dependent. Solving the
vectorial wave equation that converts the perturbations from the
Zerilli gauge to the Lorenz gauge requires numerical techniques,
except when $R \gg M$ and we can rely on approximate analytical
methods. In this section we obtain exact numerical results for
$a[l=1; \mbox{even}]$, as well as approximate analytical results for
$R \gg M$.

In Sec.~VI we discuss our results and offer a number of concluding
remarks.

\section{Newtonian self-acceleration}

In this section we consider a Newtonian system involving a large mass
$M$ and a much smaller mass $m$. The position of the small mass
relative to the center of mass is described by the vector $\bm{R}(t)$,
while the position of the larger mass is described by
$\bm{\rho}(t)$. Taking the center of mass to be at the
origin of the coordinate system, we have
\begin{equation}
m \bm{R} + M \bm{\rho} = \bm{0}.
\label{2.1}
\end{equation}
We denote the position vector of an arbitrary field point by $\bm{x}$,
and $r \equiv |\bm{x}|$ is its distance from the center of mass. We
shall also use $R \equiv |\bm{R}|$ and $\rho \equiv |\bm{\rho}|$.

\subsection*{Test-mass description}

We begin with a test-mass description of the situation, according to
which the smaller mass moves in the gravitational field of the larger
mass which is placed at the origin of the coordinate system. The
background Newtonian potential is
\begin{equation}
\Phi_0(\bm{x}) = -\frac{M}{r}
\label{2.2}
\end{equation}
and the background gravitational field is
\begin{equation}
\bm{g}_0 = -\bm{\nabla} \Phi_0 = -\frac{M}{r^3}\, \bm{x}.
\label{2.3}
\end{equation}
In this description, the smaller mass $m$ moves according to
$d^2\bm{R}/dt^2 = \bm{g}_0(\bm{x}=\bm{R})$. If the motion is circular,
then $m$ possesses a uniform angular velocity given by
\begin{equation}
{\Omega_0}^2 = \frac{M}{R^3},
\label{2.4}
\end{equation}
where $R$ is the orbital radius. These results are in close analogy
with a relativistic description in which the smaller mass is
taken to move on a geodesic of the background spacetime, in a
test-mass approximation.

\subsection*{Beyond the test-mass description: singular ``S'' and
  regular ``R'' perturbations of the Newtonian potential}

We next improve our description by incorporating the gravitational
effects produced by the smaller mass. The exact Newtonian potential is
\begin{equation}
\Phi(\bm{x}) = -\frac{M}{|\bm{x} - \bm{\rho}|}
- \frac{m}{|\bm{x} - \bm{R}|},
\label{2.5}
\end{equation}
and for $m \ll M$ this can be expressed as $\Phi(\bm{x}) =
\Phi_0(\bm{x}) + \delta \Phi(\bm{x})$, with a perturbation given by
\begin{equation}
\delta \Phi(\bm{x}) = -\frac{M}{|\bm{x} - \bm{\rho}|} + \frac{M}{r}
- \frac{m}{|\bm{x} - \bm{R}|}.
\label{2.6}
\end{equation}
This gives rise to a field perturbation $\delta \bm{g}$ that exerts a force
on the smaller mass. This is the particle's ``bare'' self-acceleration, and
the correspondence with the relativistic problem is clear.

An examination of Eq.~(\ref{2.6}) reveals that the last term on the
right-hand side diverges at the position of the smaller mass. But
since the gravitational field produced by this term is isotropic
around $\bm{R}(t)$, we know that this field will exert no force on
the particle. We conclude that the last term can be identified with
the singular ``S'' part of the perturbation,
\begin{equation}
\Phi_{\rm S}(\bm{x}) = - \frac{m}{|\bm{x} - \bm{R}|},
\label{2.7}
\end{equation}
and that the remainder makes up the regular ``R'' field,
\begin{equation}
\Phi_{\rm R}(\bm{x}) = -\frac{M}{|\bm{x} - \bm{\rho}|}
+ \frac{M}{r}.
\label{2.8}
\end{equation}
The full perturbation is then given by $\delta \Phi(\bm{x}) =
\Phi_{\rm S}(\bm{x}) + \Phi_{\rm R}(\bm{x})$, and only the ``R
potential'' affects the motion of the smaller mass. Once more the
correspondence with the relativistic problem is clear.

It is easy to check that to first order in $m/M$, Eq.~(\ref{2.8})
simplifies to
\begin{equation}
\Phi_{\rm R}(\bm{x}) = m \frac{ \bm{R} \cdot \bm{x} }{r^3};
\label{2.9}
\end{equation}
this simplification occurs because thanks to Eq.~(\ref{2.1}),
$\bm{\rho}$ is formally of order $m/M \ll 1$. The regular ``R'' part
of the field perturbation is then
\begin{equation}
\bm{g}_{\rm R}(\bm{x}) = m \frac{ 3 (\bm{R} \cdot \bm{x}) \bm{x}
- r^2 \bm{R}}{r^5},
\label{2.10}
\end{equation}
and evaluating this at the particle's position yields a correction to
the background field $\bm{g}_0(\bm{x}=\bm{R}) = -M \bm{R}/R^3$ given
by $\bm{g}_{\rm R} (\bm{x}=\bm{R}) = 2m \bm{R}/R^3$; the force still
points in the radial direction but the active mass has been shifted
from $M$ to $M-2m$. For circular motion the angular velocity becomes
\begin{equation}
\Omega^2 = \frac{M-2m}{R^3}.
\label{2.11}
\end{equation}
This can be cast in a more recognizable form if we express the
angular velocity in terms of the total separation $s \equiv R+\rho =
(1+m/M)R$ between the two masses. To first order in $m/M$ we obtain
$\Omega^2 = (M+m)/s^3$, which is just the usual form of Kepler's
third law. The regular part of the field perturbation is therefore
responsible for the finite-mass correction to the angular velocity.

\subsection*{Multipole decomposition of the perturbations}

We now examine the low-multipole content of $\delta \Phi(\bm{x})$,
$\Phi_{\rm S}(\bm{x})$, and $\Phi_{\rm R}(\bm{x})$.

It is evident from Eq.~(\ref{2.9}) that $\Phi_{\rm R}(\bm{x})$
possesses a pure dipolar form, and its multipole decomposition
therefore involves a single term at $l=1$. As we have seen, this
dipole potential is responsible for an important finite-mass
correction to the orbital frequency. We take this as a clear
indication that in the relativistic context, the $l=1$ contribution
to the metric perturbations produces important physical effects that
should not be ignored. Since there is no analogue in Newtonian theory
to the odd parity metric perturbations, this statement might be
restricted to the $l=1$, even parity perturbations of the
Schwarzschild spacetime.

While $\Phi_{\rm R}(\bm{x})$ possesses only a dipole component, the
same is not true of $\Phi_{\rm S}(\bm{x})$ and $\delta \Phi(\bm{x})$.
Their monopole components are given by
\begin{equation}
\Phi^{l=0}_{\rm S}(\bm{x}) = \delta \Phi^{l=0}(\bm{x}) =
\left\{
\begin{array}{ll}
-m/R & \qquad r < R, \\
-m/r & \qquad r > R,
\end{array} \right.
\label{2.12}
\end{equation}
and this gives rise to a monopole field perturbation
\begin{equation}
\delta \bm{g}^{l=0} =
\left\{
\begin{array}{ll}
0 & \qquad r < R, \\
-m \bm{x}/r^3 & \qquad r > R.
\end{array} \right.
\label{2.13}
\end{equation}
This is discontinuous at $\bm{x} = \bm{R}$: the field is zero when
the limit is taken from the inside, and equal to $-m \bm{R}/R^3$ when
taken from the outside. The jump in the monopole field perturbation
is given by
\begin{eqnarray}
\bigl[ \delta \bm{g}^{l=0} \bigr] &\equiv&
\delta \bm{g}^{l=0}(\bm{x} = \bm{R}) \Bigr|_{\rm outside}
- \delta \bm{g}^{l=0}(\bm{x} = \bm{R}) \Bigr|_{\rm inside}
\nonumber \\
&=& - \frac{m}{R^3} \bm{R}.
\label{2.14}
\end{eqnarray}
These results, which could be described as the Newtonian ``bare''
self-acceleration for $l=0$, will be recovered as limits of our exact
relativistic expressions in Sec.~III.

The dipole component of the singular potential is calculated to be
\begin{equation}
\Phi^{l=1}_{\rm S}(\bm{x}) =
\left\{
\begin{array}{ll}
-m (\bm{R} \cdot \bm{x})/R^3 & \qquad r < R, \\
-m (\bm{R} \cdot \bm{x})/r^3 & \qquad r > R,
\end{array} \right.
\label{2.15}
\end{equation}
and adding this to Eq.~(\ref{2.9}) we find
\begin{equation}
\delta \Phi^{l=1}(\bm{x}) = m (\bm{R} \cdot \bm{x}) \biggl(
\frac{1}{r^3} - \frac{1}{R^3} \biggr) \qquad (r < R)
\label{2.16}
\end{equation}
and $\delta \Phi^{l=1}(\bm{x}) = 0$ for $r>R$. This gives rise to the
field perturbation
\begin{equation}
\delta \bm{g}^{l=1} = m \biggl[ \frac{ 3 (\bm{R} \cdot \bm{x}) \bm{x}
- r^2 \bm{R}}{r^5} + \frac{\bm{R}}{R^3} \biggr] \qquad (r < R)
\label{2.17}
\end{equation}
and $\delta \bm{g}^{l=1} = 0$ for $r>R$. The dipole field also is
discontinuous at $\bm{x} = \bm{R}$: it is zero when the limit is taken
from the outside, and equal to $3m \bm{R}/R^3$ when taken from the
inside. Its jump, defined as in Eq.~(\ref{2.14}), is given by
\begin{equation}
\bigl[ \delta \bm{g}^{l=0} \bigr]  = - \frac{3m}{R^3} \bm{R}.
\label{2.18}
\end{equation}
These results, which could be described as the Newtonian ``bare''
self-acceleration for $l=1$, will also be recovered as limits of our
exact relativistic expressions in Sec.~V.

\section{Monopole gravitational perturbations}

Our task in this section is to calculate the $l=0$ metric
perturbations of the Schwarzschild spacetime produced by a particle of
mass $m$ in circular orbit at a radius $R$. We shall also calculate
the associated contribution to the self-acceleration, $a[l=0]$ as
defined by Eq.~(\ref{1.3}). We need the perturbations in the Lorenz
gauge, and our strategy will be to obtain them first in the simpler
Zerilli gauge, and then look for a transformation to the Lorenz gauge.

\subsection*{Perturbations in the Zerilli gauge}

The monopole perturbations produced by a point particle in arbitrary
motion around a Schwarzschild black hole were first computed by
Zerilli \cite{zerilli:70}. With his specific choice of gauge for
circular motion, the metric perturbations are
\begin{equation}
h^{\rm Z}_{tt} = 2 m \tilde{E} \Bigl( \frac{1}{r}
- \frac{f}{R-2M} \Bigr) \Theta(r-R)
\label{3.1}
\end{equation}
and
\begin{equation}
h^{\rm Z}_{rr} = \frac{2 m \tilde{E}}{r f^2} \Theta(r-R),
\label{3.2}
\end{equation}
where $f=1-2M/r$, $\tilde{E} = (1-2M/R)(1-3M/R)^{-1/2}$ is the
particle's energy per unit rest mass, and $\Theta(r-R)$ is the
Heaviside step function. It is easy to check that for $r>R$,
$g_{\alpha\beta} + h^{\rm Z}_{\alpha\beta}$ is another Schwarzschild
metric with mass parameter $M + m\tilde{E}$. The perturbation
therefore describes the sudden shift in mass parameter that occurs at
$r=R$.

\subsection*{Transformation to the Lorenz gauge}

The metric perturbation of Eqs.~(\ref{3.1}) and (\ref{3.2}) does not
satisfy the Lorenz gauge condition of Eq.~(\ref{1.2}). We therefore
seek a vector field $\xi^\alpha$ that generates a transformation from
the Zerilli gauge to the Lorenz gauge. This vector must possess only
an $l=0$ component, and so it must be of the form $\xi_\alpha = [0,
\xi(r), 0, 0]$. As the perturbation is static, there is no need to
include a component in the time direction (this point is elaborated in
the Appendix), nor a time dependence in the radial component.

To find this vector we express the Lorenz-gauge metric perturbations
in the standard Regge-Wheeler \cite{regge-wheeler:57} form
\begin{eqnarray}
h_{tt} &=& f H_0(r), \nonumber \\
h_{rr} &=& H_2(r)/f, \label{3.3} \\
h_{AB} &=& r^2 \Omega_{AB} K(r), \nonumber
\end{eqnarray}
where the upper-case latin indices run over the angular coordinates
$\theta$ and $\phi$, and $\Omega_{AB} = \mbox{diag}(1,\sin^2\theta)$
is the metric of the unit two-sphere. We have set $h_{tr} = H_1(r) =
0$ on the grounds that the perturbation must be static. A similar
notation can be used to express the Zerilli-gauge perturbations, and
we have $K^{\rm Z} = 0$ while $H_0^{\rm Z}$ and $H_2^{\rm Z}$ are
nonzero. The gauge transformation is given by
$h_{\alpha\beta} = h^{\rm Z}_{\alpha\beta}
- 2\xi_{(\alpha;\beta)}$ and this translates to
\begin{eqnarray}
H_0 &=& H^{\rm Z}_0 + \frac{2M}{r^2} \xi, \nonumber \\
H_2 &=& H^{\rm Z}_2 - 2f \xi' - \frac{2M}{r^2} \xi, \label{3.4} \\
K &=& -\frac{2f}{r} \xi, \nonumber
\end{eqnarray}
where a prime indicates differentiation with respect to $r$. The new
perturbation will satisfy the Lorenz gauge condition if
\begin{equation}
f\bigl( H_0' + H_2' - 2K' \bigr) + \frac{2M}{r^2} H_0
+ \frac{2(2r-3M)}{r^2} H_2 - \frac{4f}{r} K = 0.
\label{3.5}
\end{equation}
Using Eqs.~(\ref{3.1}), (\ref{3.2}), and (\ref{3.4}), this becomes an
ordinary differential equation for $\xi(r)$:
\begin{eqnarray}
\hspace*{-15pt}
f \xi'' + \frac{2}{r} \xi' - \frac{2f}{r^2} \xi &=&
\frac{m\tilde{E}}{R-2M} \delta(r-R)
\nonumber \\ & & \mbox{}
+ \frac{2m\tilde{E}}{r^2 f} \frac{R-3M}{R-2M} \Theta(r-R).
\label{3.6}
\end{eqnarray}
Our task is now to find a solution to this equation.

The function $\xi(r)$ can be expressed as a superposition of interior
and exterior solutions,
\begin{equation}
\xi(r) = \xi_<(r) \Theta(R-r) + \xi_>(r) \Theta(r-R).
\label{3.7}
\end{equation}
The interior solution $\xi_<(r)$ satisfies the homogeneous version of
Eq.~(\ref{3.6}), while the exterior solution $\xi_>(r)$ satisfies
Eq.~(\ref{3.6}) with $\delta(r-R)$ set equal to zero and $\Theta(r-R)$
set equal to 1. The solutions must comply with the jump conditions
\begin{equation}
\bigl[ \xi \bigr] = 0, \qquad
\bigl[ \xi' \bigr] = \frac{m \tilde{E} R}{(R-2M)^2},
\label{3.8}
\end{equation}
where $[\xi] \equiv \xi_>(r=R) - \xi_<(r=R)$ and a similar definition
holds for $[\xi']$. Equations (\ref{3.8}) and (\ref{3.4}) imply that
in the Lorenz gauge, the metric perturbations are continuous at $r=R$:
$[H_0] = [H_2] = [K] = 0$.

The interior solution is a linear superposition of the two independent
solutions $\xi_1 = [r(r-2M)]^{-1}$ and $\xi_2 =
r^2/(r-2M)$. Regularity at the event horizon requires that $\xi$ be
well behaved in the limit $r \to 2M$. We must therefore choose
\begin{equation}
\xi_<(r) = a \frac{r^2 + 2Mr + 4M^2}{r},
\label{3.9}
\end{equation}
where $a$ is a constant that will be determined by the jump
conditions. The exterior solution is a linear superposition of
$\xi_1$, $\xi_2$, and the particular solution $\xi_p = - m\tilde{E}
(R-3M)(R-2M)^{-1}\Gamma(r)$, where
\begin{eqnarray}
\Gamma(r) &=& - \Bigl\{ [9 M r(r-2M) \bigr]^{-1} \bigl[ 3r^3
\ln(1-2M/r) - 3Mr^2 \nonumber \\ & & \hspace*{-15pt} \mbox{} - 12M^2
r + 44M^3 - 24M^3 \ln\bigl[r/(2M) - 1\bigr] \Bigr\}. \label{3.10}
\end{eqnarray}
Because $\xi_1 \sim 1/r^2$, $\xi_2 \sim r$, and $\Gamma \sim 1$ when
$r\to\infty$, proper asymptotic behavior requires that we discard
$\xi_2$ from the exterior solution. We then have
\begin{equation}
\xi_>(r) = b \frac{M^3}{r(r-2M)}
- m\tilde{E} \frac{R-3M}{R-2M} \Gamma(r),
\label{3.11}
\end{equation}
where $b$ is a constant that will be determined by the jump
conditions.

The gauge vector is now fully determined: The interior solution is
given by Eq.~(\ref{3.9}) and the exterior solution by Eq.~(\ref{3.11})
with the function $\Gamma(r)$ displayed in Eq.~(\ref{3.10}). The
complete gauge vector field is then constructed as in Eq.~(\ref{3.7}),
and the constants $a$ and $b$ are determined by the jump conditions of
Eq.~(\ref{3.8}). This is sufficient information to calculate the
Lorenz-gauge metric perturbations with the help of Eqs.~(\ref{3.3})
and (\ref{3.4}). Because the resulting expressions are moderately
lengthy, we shall not display these results here, but proceed instead
with the calculation of the self-acceleration.

Before moving on we wish to call attention to the fact that in the
foregoing manipulations, the requirements of staticity, regularity at
the event horizon, and regularity at infinity have allowed us to
construct a {\it unique solution} to the perturbation equations in the
Lorenz gauge. This conclusion is elaborated in the Appendix.

\subsection*{Monopole contribution to the self-acceleration}

The self-acceleration produced by the $l=0$ perturbations can be
expressed as a sum of two terms,
\begin{equation}
a[l=0] = a[l=0; \mbox{Zerilli}] + a[l=0; \mbox{gauge}],
\label{3.12}
\end{equation}
where $a[l=0; \mbox{Zerilli}]$ is the radial component of the
acceleration vector constructed as in Eq.~(\ref{1.1}) but by
replacing $h^{\rm R}_{\alpha\beta}$ with $h^{\rm Z}_{\alpha\beta}$,
while $a[l=0; \mbox{gauge}]$ is constructed from $h^{\rm
gauge}_{\alpha\beta} = -2\xi_{(\alpha;\beta)}$. The calculation
involves the particle's velocity vector $u^\mu =
(1-3M/R)^{-1/2}[1,0,0,\Omega]$, and at the end
$h_{\alpha\beta;\gamma}$ must be evaluated at the position of the
particle ($r=R$, $\theta=\pi/2$, and $\phi = \Omega t$), either from
the orbit's interior ($r < R$) or from its exterior ($r>R$). This
leads to two different values for the acceleration, $a_<$ and $a_>$,
respectively. Such a discontinuity was encountered before in a
Newtonian context --- refer back to Eq.~(\ref{2.13}).

The external value of the Zerilli acceleration is given by
\begin{equation}
a_>[l=0; \mbox{Zerilli}] = -\frac{m\tilde{E}}{R(R-3M)},
\label{3.13}
\end{equation}
while the internal value is zero: $a_<[l=0; \mbox{Zerilli}] = 0$.
The gauge acceleration, on the other hand, is found to be
\[
a[l=0; \mbox{gauge}] = -\frac{3M(R-2M)^2}{R^4(R-3M)} \xi(R),
\]
and by virtue of Eq.~(\ref{3.8}), the internal and external values are
equal. The gauge vector can most simply be evaluated from the orbit's
interior, and Eq.~(\ref{3.9}) gives $\xi(R) = a (R^2 + 2MR +
4M^2)/R$. But the jump conditions imply $a = \frac{1}{3}
(m \tilde{E}/M) [(R-3M)\ln(1-2M/R) - M]/(R-2M)$, and altogether we
obtain
\begin{eqnarray}
\hspace*{-5pt}
a[l=0; \mbox{gauge}] &=& m\tilde{E}
\frac{ (R-2M)(R^2 + 2MR + 4M^2) }{R^5}
\nonumber \\ & & \mbox{} \times
\biggl[ \frac{M}{R-3M} - \ln \biggl(1 - \frac{2M}{R} \biggr) \biggr].
\label{3.14}
\end{eqnarray}

From Eqs.~(\ref{3.12})--(\ref{3.14}) we arrive at our final
results. The internal value for the $l=0$ self-acceleration is
\begin{eqnarray}
a_<[l=0] &=& m\tilde{E}
\frac{ (R-2M)(R^2 + 2MR + 4M^2) }{R^5}
\nonumber \\ & & \mbox{} \times
\biggl[ \frac{M}{R-3M} - \ln \biggl(1 - \frac{2M}{R} \biggr) \biggr],
\label{3.15}
\end{eqnarray}
while the external value is
\begin{eqnarray}
a_>[l=0] &=& -m\tilde{E} \frac{R^4 - MR^3 + 8M^4}{R^5 (R-3M)}
\nonumber \\ & & \mbox{}
- m\tilde{E} \frac{ (R-2M)(R^2 + 2MR + 4M^2) }{R^5}
\nonumber \\ & & \hspace*{20pt} \mbox{} \times
\ln \biggl(1 - \frac{2M}{R} \biggr).
\label{3.16}
\end{eqnarray}
When $R \gg M$ these expressions simplify to $a_<[l=0] \sim 3 m
M/R^3$ and $a_>[l=0] \sim -m/R^2 + mM/(2R^3)$; the internal value is
smaller than the external value by a factor of order $M/R \ll
1$. These limiting expressions are compatible with the Newtonian
results displayed in Eq.~(\ref{2.13}). They differ, however, from
the results of Nakano, Sago, and Sasaki \cite{nakano-etal:03}, which
are displayed in their Eq.~(E19) --- our expressions are smaller than
theirs by a term $4mM/R^3$. This discrepancy is explained in the
Appendix. Equations (\ref{3.15}) and (\ref{3.16}) were used to
generate the curves shown in Figs.~1 and 2.

\section{Dipole, odd-parity gravitational perturbations}

In this section we calculate the $l=1$, odd-parity metric
perturbations of the Schwarzschild spacetime produced by a particle
of mass $m$ in circular orbit at a radius $R$. From these we shall
derive their contribution to the self-acceleration, $a[l=1;
\mbox{odd}]$ as defined by Eq.~(\ref{1.3}). Here we shall find that
the expressions provided by Zerilli already satisfy the Lorenz gauge
condition.

The dipole, odd-parity perturbations produced by a point particle in
arbitrary motion around a Schwarzschild black hole were first computed
by Zerilli \cite{zerilli:70} and shown to be intimately related to the
shift in angular-momentum parameter that occurs at the orbit. After
specializing to circular motion in the equatorial plane, his results
read
\begin{equation}
h_{t\phi} = -2 m \tilde{L} \sin^2\theta \times \left\{
\begin{array}{ll}
r^2/R^3 & \qquad r < R, \\
1/r     & \qquad r > R,
\end{array} \right.
\label{4.1}
\end{equation}
where $\tilde{L} = [M R/(1-3M/R)]^{1/2}$ is the particle's angular
momentum per unit rest mass. For $r<R$, the metric $g_{\alpha\beta} +
h_{\alpha\beta}$ differs from $g_{\alpha\beta}$ only by a gauge
transformation --- it is also a Schwarzschild metric with mass
parameter $M$. For $r>R$, $g_{\alpha\beta} + h_{\alpha\beta}$ is a
Kerr metric linearized with respect to the angular-momentum
parameter $a \equiv (m/M)\tilde{L}$. The perturbation therefore
describes the sudden shift in angular momentum that occurs at $r=R$.

It is easy to check that the perturbation of Eq.~(\ref{4.1})
satisfies the Lorenz gauge condition of Eq.~(\ref{1.2}). It is also
easy to show that a (time-independent) gauge transformation within
the class of Lorenz gauges would produce a pathological behavior of
the perturbation at the event horizon. Equation (\ref{4.1}) therefore
gives us a unique solution to the perturbation equations in the
Lorenz gauge.

A straightforward calculation then reveals that the internal value of
the $l=1$, odd-parity contribution to the self-acceleration is
\begin{equation}
a_<[l=1;\mbox{odd}] = - \frac{4 m M}{R^3}
\frac{1-2M/R}{(1-3M/R)^{3/2}},
\label{4.2}
\end{equation}
while the external value is
\begin{equation}
a_>[l=1;\mbox{odd}] = \frac{2 m M}{R^3}
\frac{1-2M/R}{(1-3M/R)^{3/2}}.
\label{4.3}
\end{equation}
These results have no analogue in Newtonian theory. Equations
(\ref{4.2}) and (\ref{4.3}) were used to generate the curves shown in
Figs.~1 and 2.

\section{Dipole, even-parity gravitational perturbations}

Our task in this section is to calculate the $l=1$, even-parity metric
perturbations of the Schwarzschild spacetime produced by a particle of
mass $m$ in circular orbit at a radius $R$. We shall also calculate
the associated contribution to the self-acceleration,
$a[l=1; \mbox{even}]$ as defined by Eq.~(\ref{1.3}). Once more we need
the perturbations in the Lorenz gauge, and as in Sec.~III our strategy
will be to obtain them first in the simpler Zerilli gauge, and then
look for a transformation to the Lorenz gauge. The solution to the
wave equation satisfied by the gauge vector field will be obtained
numerically and provided in tabulated form. It will also be obtained
analytically in a post-Newtonian expansion in powers of $M/R$.

\subsection*{Perturbations in the Zerilli gauge}

The dipole, even-parity perturbations produced by a point particle in
arbitrary motion around a Schwarzschild black hole were first computed
by Zerilli \cite{zerilli:70} in a simple choice of gauge. After
specializing to circular motion, his results become
\begin{eqnarray}
h^{\rm Z}_{tt} &=& 2 m \tilde{E}\, \frac{R-2M}{r(r-2M)} \bigl( 1 - r^3
\Omega^2/M \bigr)
\nonumber \\ & & \times
\sin\theta \cos(\phi - \Omega t) \Theta(r-R),
\label{5.1} \\
h^{\rm Z}_{tr} &=& -6 m \tilde{E}\, \Omega (R-2M) \frac{r}{(r-2M)^2}
\nonumber \\ & & \times
\sin\theta \sin(\phi - \Omega t) \Theta(r-R),
\label{5.2} \\
h^{\rm Z}_{rr} &=& 6 m \tilde{E}\, (R-2M) \frac{r}{(r-2M)^3}
\nonumber \\ & & \times
\sin\theta \cos(\phi - \Omega t) \Theta(r-R),
\label{5.3}
\end{eqnarray}
where $\Omega = \sqrt{M/R^3}$ is the particle's angular velocity and
$\tilde{E} = (1-2M/R)(1-3M/R)^{-1/2}$ is its energy per unit
mass. Here we see that the perturbations are time dependent, and this
complicates considerably the task of finding the transformation to the
Lorenz gauge. Equation (\ref{5.1}) reveals that the Zerilli gauge is
not asymptotically flat, since $h^{\rm Z}_{tt}$ grows linearly with
$r$ as $r \to \infty$. This indicates the fact that the metric
$g_{\alpha\beta} + h^{\rm Z}_{\alpha\beta}$ is expressed in a
noninertial coordinate system anchored to the black hole instead of
the system's center of mass. This statement will be elaborated below.

\subsection*{Perturbations in a singular gauge}

The metric perturbations of Eqs.~(\ref{5.1})--(\ref{5.3}) do not
satisfy the Lorenz gauge condition of Eq.~(\ref{1.2}). To transform to
the Lorenz gauge we proceed in two steps. We shall first transform to
a gauge in which the perturbation is zero everywhere, except at $r=R$
where it is singular. We shall then go from this singular gauge
to the Lorenz gauge.

It is well known from Zerilli's work \cite{zerilli:70} that {\it in
vacuum}, a dipole, even-parity perturbation can be completely removed
by a gauge transformation. Such a perturbation, therefore, represents
a coordinate transformation; and as we have already suggested, for
$r>R$ the metric $g_{\alpha\beta} + h^{\rm Z}_{\alpha\beta}$ is just
a Schwarzschild solution expressed in a noninertial coordinate
system. The perturbations of Eqs.~(\ref{5.1})--(\ref{5.3}), however,
are not pure gauge because of the presence of the particle. They can
be removed in the vacuum region outside of $r=R$, but the gauge
transformation leaves something behind at $r=R$. The result of this
transformation is $h^{\rm s}_{\alpha\beta}$, the metric perturbation
in what we shall call the {\it singular gauge}.

The gauge transformation that removes a dipole, even-parity
perturbation in vacuum was constructed by Zerilli \cite{zerilli:70}.
It is generated by a vector field $\varepsilon^\alpha$, so that
$h^{\rm s}_{\alpha\beta} = h^{\rm Z}_{\alpha\beta} -
2\varepsilon_{(\alpha;\beta)}$. For circular motion this is given by
\begin{eqnarray}
\varepsilon_t &=& \frac{m\tilde{E}}{M}\, \Omega (R-2M) \frac{r^2}{r-2M}
\nonumber \\ & & \times
\sin\theta \sin(\phi - \Omega t)\, \Theta(r-R),
\label{5.4} \\
\varepsilon_r &=& -\frac{m\tilde{E}}{M}\, (R-2M) \frac{r^2}{(r-2M)^2}
\nonumber \\ & & \times
\sin\theta \cos(\phi - \Omega t)\, \Theta(r-R),
\label{5.5} \\
\varepsilon_\theta &=& -\frac{m\tilde{E}}{M}\, (R-2M) \frac{r^2}{r-2M}
\nonumber \\ & & \times
\cos\theta \cos(\phi - \Omega t)\, \Theta(r-R),
\label{5.6} \\
\varepsilon_\phi &=& \frac{m\tilde{E}}{M}\, (R-2M) \frac{r^2}{r-2M}
\nonumber \\ & & \times
\sin\theta \sin(\phi - \Omega t)\, \Theta(r-R).
\label{5.7}
\end{eqnarray}
The new metric perturbation is then
\begin{eqnarray}
\hspace*{-13pt} h^{\rm s}_{tr} &=& -\frac{m \tilde{E}}{M}\, \Omega
R^2 \sin\theta \sin(\phi - \Omega t)\, \delta(r-R),
\label{5.8} \\
\hspace*{-13pt} h^{\rm s}_{rr} &=& 2 \frac{m \tilde{E}}{M}\,
\frac{R^2}{R-2M} \sin\theta \cos(\phi - \Omega t)\, \delta(r-R),
\label{5.9} \\
\hspace*{-13pt} h^{\rm s}_{r\theta} &=& \frac{m \tilde{E}}{M}\, R^2
\cos\theta \cos(\phi - \Omega t)\, \delta(r-R),
\label{5.10} \\
\hspace*{-13pt} h^{\rm s}_{r\phi} &=& -\frac{m \tilde{E}}{M}\, R^2
\sin\theta \sin(\phi - \Omega t)\, \delta(r-R), \label{5.11}
\end{eqnarray}
and we see that in the singular gauge, the metric perturbation is
proportional to $\delta(r-R)$, which is produced by differentiation
of the step function in $\varepsilon_\alpha$. The gauge
transformation therefore makes the perturbation zero everywhere in
the vacuum region outside (and inside) $r=R$, but it contributes a
singular term at the orbit. This illustrates the fact that the
presence of matter prevents the Zerilli-gauge metric perturbation
from being pure gauge.

\subsection*{Interpretation of the gauge transformation}

The preceding discussion on coordinate systems can be clarified if we
examine the asymptotic behavior of the gauge vector field in the
limit $r \to \infty$. In this limit we can seek a Newtonian
interpretation of the results, and we shall see that in the original
Zerilli gauge, the perturbed metric is that of a moving black
hole. The following is patterned after a similar discussion produced
by Zerilli \cite{zerilli:70}.

The vector $\varepsilon^\alpha$ becomes asymptotically equal to
$b^\alpha$ in the limit $r \to \infty$, where
\begin{eqnarray*}
b^t &=& -\frac{m \tilde{E}}{M} (R-2M) \frac{\partial}{\partial t}\,
r \sin\theta \cos(\phi - \Omega t), \\
b^r &=& -\frac{m \tilde{E}}{M} (R-2M)
\sin\theta \cos(\phi - \Omega t), \\
b^\theta &=& -\frac{m \tilde{E}}{M} (R-2M)
\frac{1}{r} \cos\theta \cos(\phi - \Omega t), \\
b^\phi &=& \frac{m \tilde{E}}{M} (R-2M)
\frac{1}{r\sin\theta} \sin(\phi - \Omega t).
\end{eqnarray*}
If we introduce asymptotic Cartesian coordinates $x=r\sin\theta
\cos\phi$, $y=r\sin\theta \sin\phi$, and $z=r\cos\theta$, we have
\begin{eqnarray*}
b^t &=& -\frac{m \tilde{E}}{M} (R-2M) \frac{\partial}{\partial t}
(x \cos\Omega t + y \sin\Omega t), \\
b^x &=& -\frac{m \tilde{E}}{M} (R-2M) \cos\Omega t, \\
b^y &=& -\frac{m \tilde{E}}{M} (R-2M) \sin\Omega t, \\
b^z &=& 0.
\end{eqnarray*}

To give a Newtonian interpretation to these results, let $\bm{x}$ be
the Zerilli coordinates of an arbitrary field point, let $\bm{R}(t)
\equiv (R-2M)(\cos\Omega t, \sin\Omega t, 0)$ be the position vector
of the orbiting particle, and express the preceding equations as
\[
\bm{b}(t) = - \frac{ m\tilde{E} }{M} \bm{R}(t), \qquad
b^t = \bm{x} \cdot \dot{\bm{b}}(t),
\]
where an overdot indicates differentiation with respect to $t$. The
coordinate transformation generated by $b^\alpha$ is then
\[
\bm{x}_{\rm new} = \bm{x} + \bm{b}(t), \qquad
t_{\rm new} = t + \bm{x} \cdot \dot{\bm{b}}(t).
\]
We can now explain that this transformation represents a translation
from a noninertial reference frame attached to the black hole to an
inertial frame attached to the center of mass. Please refer
back to Sec.~II for a definition of the notation employed here.

In the center-of-mass frame, the particle moves on a trajectory
$\bm{R}_{\rm cm}(t)$, and the black hole moves on a trajectory
$\bm{\rho}_{\rm cm}(t)$. In the black-hole frame we have
$\bm{R}_{\rm bh}(t) \equiv \bm{R}(t)$ and
$\bm{\rho}_{\rm bh}(t) \equiv \bm{0}$. The center-of-mass condition is
$(m\tilde{E}) \bm{R}_{\rm cm} + M \bm{\rho}_{\rm cm} = 0$, so we have
$\bm{\rho}_{\rm cm} = -(m\tilde{E}/M) \bm{R}_{\rm cm}$. We also have
$\bm{R} = \bm{R}_{\rm cm} - \bm{\rho}_{\rm cm} = (1+m\tilde{E}/M)
\bm{R}_{\rm cm}$, so that $\bm{R}_{\rm cm} = \bm{R} + O(m/M)$ and
$\bm{\rho}_{\rm cm} = -(m \tilde{E} /M) \bm{R} + O(m^2/M^2)$, or
\[
\bm{\rho}_{\rm cm}(t) = \bm{b}(t) + O(m^2/M^2).
\]
The vector $\bm{b}(t)$ is therefore the position of the black hole
relative to the center of mass, and the coordinate transformation is
truly a translation from the moving frame of the black hole
to the fixed reference frame of the center of mass.

\subsection*{Transformation to the Lorenz gauge}

We now return to the task of transforming the metric perturbation from
the singular gauge of Eqs.~(\ref{5.8})--(\ref{5.11}) to the Lorenz
gauge. The gauge transformation is generated by a vector field
$\xi^\alpha$, such that
\begin{equation}
h_{\alpha\beta} = h^{\rm s}_{\alpha\beta} - 2\xi_{(\alpha;\beta)}
\label{5.12}
\end{equation}
is the Lorenz-gauge metric perturbation. For this to comply with
Eq.~(\ref{1.2}), the vector field must satisfy the inhomogeneous wave
equation
\begin{equation}
\Box \xi^\alpha = S^\alpha,
\label{5.13}
\end{equation}
where $\Box = \nabla^\beta \nabla_\beta$ is the wave operator and
\begin{equation}
S^\alpha = \nabla_\beta \Bigl( h_{\rm s}^{\alpha\beta} - \frac{1}{2}
g^{\alpha\beta} g^{\gamma\delta} h^{\rm s}_{\gamma\delta} \Bigr)
\label{5.14}
\end{equation}
is the source term. This is given explicitly by
\begin{eqnarray}
\hspace*{-20pt}
S^t &=& \frac{m \tilde{E}}{M} \frac{\Omega R}{R-2M} \Bigl[
(3R-2M) \delta(r-R)
\nonumber \\ & & \hspace*{-8pt} \mbox{}
+ R(R-2M) \delta'(r-R) \Bigr]
\sin\theta \sin(\phi - \Omega t),
\label{5.15} \\
\hspace*{-20pt}
S^r &=& \frac{m \tilde{E}}{M} \frac{1}{R} \Bigl[
(2R-5M) \delta(r-R)
\nonumber \\ & & \hspace*{-8pt} \mbox{}
+ R(R-2M) \delta'(r-R) \Bigr]
\sin\theta \cos(\phi - \Omega t),
\label{5.16} \\
\hspace*{-20pt}
S^\theta &=& \frac{m \tilde{E}}{M} \frac{1}{R^2} \Bigl[
(3R-8M) \delta(r-R)
\nonumber \\ & & \hspace*{-8pt} \mbox{}
+ R(R-2M) \delta'(r-R) \Bigr]
\cos\theta \cos(\phi - \Omega t),
\label{5.17} \\
\hspace*{-20pt}
S^\phi &=& -\frac{m \tilde{E}}{M} \frac{1}{R^2 \sin^2\theta} \Bigl[
(3R-8M) \delta(r-R)
\nonumber \\ & & \hspace*{-8pt} \mbox{}
+ R(R-2M) \delta'(r-R) \Bigr]
\sin\theta \sin(\phi - \Omega t).
\label{5.18}
\end{eqnarray}
To arrive at these results we have invoked the
distributional identity $g(r) \delta'(r-R) = g(R) \delta'(r-R) -
g'(R)\delta(r-R)$, where $g(r)$ is any test function and $g'(r)$ its
derivative with respect to $r$.

To solve Eq.~(\ref{5.13}) we decompose the vector $\xi_\alpha$ in
even-parity spherical harmonics of degree $l=1$. The form of the
source term indicates that only terms with $m = \pm 1$ are needed, and
we let
\begin{eqnarray}
\xi_a(t,r,\theta^A) &=& \sum_\pm \xi^\pm_a(t,r) Y^\pm(\theta^A),
\label{5.19} \\
\xi_A(t,r,\theta^A) &=& \sum_\pm \xi^\pm(t,r) \partial_A
Y^\pm(\theta^A).
\label{5.20}
\end{eqnarray}
Here, the lower-case latin index $a$ refers to the $t$ and $r$
components of the vector field, while the upper-case index $A$ refers
to the angular components; we have set $\theta^A = (\theta,\phi)$ and
$Y^\pm(\theta^A) \equiv Y_{l=1}^{m=\pm 1}(\theta^A) = \mp
\sqrt{3/(8\pi)} \sin\theta e^{\pm i \phi}$.

The vector $S^\alpha$ can be decomposed in a similar way, and
to simplify the form of the reduced wave equation we define the
functions $A^\pm(r)$, $B^\pm(r)$, and $C^\pm(r)$ by the relations
\begin{eqnarray}
\xi_t^\pm(t,r) &=& -\frac{1}{2} \sqrt{\frac{8\pi}{3}}
\frac{m \tilde{E}}{M} i\Omega R \frac{A^\pm(r)}{r} e^{\mp i\Omega t},
\label{5.21} \\
\xi_r^\pm(t,r) &=& \mp \frac{1}{2} \sqrt{\frac{8\pi}{3}}
\frac{m \tilde{E}}{M} \frac{B^\pm(r)}{r-2M} e^{\mp i\Omega t},
\label{5.22} \\
\xi^\pm(t,r) &=& \mp \frac{1}{2} \sqrt{\frac{8\pi}{3}}
\frac{m \tilde{E}}{M} C^\pm(r) e^{\mp i\Omega t}.
\label{5.23}
\end{eqnarray}
With these definitions Eq.~(\ref{5.13}) becomes the following set of
ordinary differential equations:
\begin{widetext}
\begin{equation}
\frac{d^2 A^\pm}{dr^2}
+ \biggl[ \frac{\Omega^2 r^2}{(r-2M)^2} - \frac{2}{r(r-2M)} \biggr]
A^\pm - \frac{2M/R}{(r-2M)^2}\, B^\pm = \frac{2R^2}{R-2M}\,
\delta(r-R) + R^2 \delta'(r-R),
\label{5.24}
\end{equation}
\begin{equation}
\frac{d^2 B^\pm}{dr^2} + \biggl[ \frac{\Omega^2 r^2}{(r-2M)^2} -
\frac{4(r-M)}{r^2(r-2M)} \biggr] B^\pm + \frac{2\Omega^2MR}{(r-2M)^2}\, A^\pm
+ \frac{4}{r^2}\, C^\pm = \frac{R(R-M)}{R-2M}\, \delta(r-R) + R^2
\delta'(r-R), \label{5.25}
\end{equation}
\begin{equation}
 \frac{d^2C^\pm}{dr^2} + \frac{2M}{r(r-2M)}\frac{dC^\pm}{dr}
+ \biggl[ \frac{\Omega^2 r^2}{(r-2M)^2} - \frac{2}{r(r-2M)} \biggr]
C^\pm + \frac{2}{r(r-2M)}\, B^\pm = R\, \delta(r-R)
+ R^2 \delta'(r-R).
\label{5.26}
\end{equation}
\end{widetext}
The steps required to obtain the Lorenz-gauge metric perturbation are
therefore these: First, solve Eqs.~(\ref{5.24})--(\ref{5.26}) for the
functions $A^\pm(r)$, $B^\pm(r)$, and $C^\pm(r)$; second, insert the
solutions into Eqs.~(\ref{5.21})--(\ref{5.23}), and these into
Eqs.~(\ref{5.19}) and (\ref{5.20}), to construct the gauge vector
field $\xi_\alpha$; third, compute $h_{\alpha\beta}$ using
Eq.~(\ref{5.12}).

\subsection*{Jump conditions and asymptotic behavior}

The solutions to Eqs.~(\ref{5.24})--(\ref{5.26}) can be expressed as
\begin{eqnarray}
A^\pm(r) &=& A^\pm_<(r) \Theta(R-r) + A^\pm_>(r) \Theta(r-R),
\label{5.27} \\
B^\pm(r) &=& B^\pm_<(r) \Theta(R-r) + B^\pm_>(r) \Theta(r-R),
\label{5.28} \\
C^\pm(r) &=& C^\pm_<(r) \Theta(R-r) + C^\pm_>(r) \Theta(r-R),
\label{5.29}
\end{eqnarray}
where the interior and exterior solutions satisfy the corresponding
homogeneous equations. To account for the source terms, these
functions must comply with the jump conditions
\begin{equation}
\bigl[A^\pm \bigr] = \bigl[B^\pm \bigr] = \bigl[C^\pm \bigr] = R^2
\label{5.30}
\end{equation}
and
\begin{eqnarray}
\biggl[ \frac{dA^\pm}{dr} \biggr] &=& \frac{2R^2}{R-2M},
\label{5.31} \\
\biggl[ \frac{dB^\pm}{dr} \biggr] &=& \frac{R(R-M)}{R-2M},
\label{5.32} \\
\biggl[ \frac{dC^\pm}{dr} \biggr] &=& \frac{R(R-4M)}{R-2M},
\label{5.33}
\end{eqnarray}
where $[\psi] \equiv \psi_>(r=R) - \psi_<(r=R)$ for any function
$\psi$ of the radial coordinate.

Near the event horizon the interior functions can be expanded as
\begin{eqnarray}
 A_<^\pm(r) &=& e^{\mp i\Omega r^*} \sum_{n=0}^\infty a_n(r-2M)^n,
\label{5.34} \\
 B_<^\pm(r) &=& e^{\mp i\Omega r^*} \sum_{n=0}^\infty b_n(r-2M)^n,
\label{5.35} \\
 C_<^\pm(r) &=& e^{\mp i\Omega r^*} \sum_{n=0}^\infty c_n(r-2M)^n .
\label{5.36}
\end{eqnarray}
These forms ensure that the vector $\xi_\alpha$ satisfies ingoing-wave
boundary conditions at the horizon, a necessary condition to obtain a
{\it retarded solution} to Eq.~(\ref{5.13}). Substitution into
Eqs.~(\ref{5.24})--(\ref{5.26}) provides recurrence relations
consisting of three coupled expressions for the $a_n$, $b_n$, and
$c_n$:
\begin{widetext}
\begin{eqnarray}
 [ 2M \lefteqn{ n( n-1) \mp 4i \Omega M^2 (2n-1) ]  a_n
   -  4M^2 R^{-1} b_n } &&
\nonumber\\ && =
   - [  n(n-3) \mp 2i\Omega M (4n -5)]  a_{n-1}
  \mp  2i\Omega (2-n) a_{n-2} + 2 M R^{-1} b_{n-1},
\label{rrAL}
\end{eqnarray}
\begin{eqnarray}
  8 \Omega^2
  \lefteqn{M^3 R a_n  + [4 M^2 n(n-1) \mp 8i\Omega M^3 (2n-1) ] b_n
  } &&
\nonumber\\ & = &
  - 8\Omega^2 M^2 R a_{n-1}
  - 2 \Omega^2 M R a_{n-2}
  - [ 4 M (n^2-3n+1) \mp 8i\Omega M^2 (3n-4)] b_{n-1}
\nonumber\\ && {}
  - [ n^2 - 5n + 2 \mp 2i\Omega M (6n-13) ] b_{n-2}
  \mp 2 i\Omega (3-n) b_{n-3}
  - 4 c_{n-2},
\label{rrBL}
\end{eqnarray}
\begin{eqnarray}
 \hskip-20pt  2 n M (n \mp 4i\Omega M) c_{n} &=&
  - 2b_{n-1}
  - [n(n-3) \mp 8i\Omega M (n-1) ]  c_{n-1}
  \mp 2i\Omega(n-2) c_{n-2}.
\label{rrCL}
\end{eqnarray}
\end{widetext}
For $n<0$, $a_n$, $b_n$, and $c_n$ are zero. For $n=0$ and 1,
Eqs.~(\ref{rrAL})--(\ref{rrCL}) allow $a_0$, $a_1$, and $c_0$ to be
chosen freely. Other early coefficients in the sequences are
\begin{eqnarray}
b_0 &=& \pm i\Omega R a_0, \\
b_1 &=& - \frac{R a_0}{2 M^2} \mp i \Omega R a_1, \\
c_1 &=& \frac{c_0 \mp i\Omega R a_1}{M(1 \mp 4i\Omega M)}.
\end{eqnarray}

Similarly, for large $r$ the exterior functions can be expanded as
\begin{eqnarray}
A_>^\pm(r) &=& e^{\pm i\Omega r^*} \sum_{n=0}^\infty \hat{a}_n r^{-n} ,
 \label{5.40} \\
B_>^\pm(r) &=& e^{\pm i\Omega r^*} \sum_{n=0}^\infty \hat{b}_n r^{-n} ,
 \label{5.41} \\
C_>^\pm(r) &=& e^{\pm i\Omega r^*} \sum_{n=0}^\infty \hat{c}_n r^{-n},
 \label{5.42}
\end{eqnarray}
These forms ensure that the vector $\xi_\alpha$ satisfies
outgoing-wave boundary conditions at infinity, another necessary
condition to obtain a {\it retarded solution} to Eq.~(\ref{5.13}).
Substitution into Eqs.~(\ref{5.24})--(\ref{5.26}) provides
recurrence relations consisting of three coupled expressions for the
$\hat{a}_n$, $\hat{b}_n$, and $\hat{c}_n$:
\begin{widetext}
\begin{eqnarray}
 \pm 2i\Omega n \hat{a}_n &=& [(n-2)(n+1) \pm 2i\Omega M (2n-3)] \hat{a}_{n-1}
  -  4M(n^2-3n+1) \hat{a}_{n-2}
\nonumber\\ && {}
  +  4M^2(n-2)(n-3) \hat{a}_{n-3}
  - 2M R^{-1} \hat{b}_{n-1},
\label{rrAR}
\end{eqnarray}
\begin{eqnarray}
  \pm 2i\Omega n \hat{b}_n &=&   2\Omega^2 R M  \hat{a}_{n-1}
  + [n^2 - n - 4 \pm 2i\Omega M(2n-3)]  \hat{b}_{n-1}
  - 4M (n^2 - 3n -1)  \hat{b}_{n-2}
\nonumber\\ && {}
  + 4M^2 (n^2 - 5n + 4) \hat{b}_{n-3}
  + 4 \hat{c}_{n-1}
  - 16 M \hat{c}_{n-2}
  +  16 M^2  \hat{c}_{m-3},
\label{rrBR}
\end{eqnarray} and
\begin{eqnarray}
   \pm 2i\Omega n \hat{c}_n &=&  2 \hat{b}_{n-1}
    + (n-2)(n+1) \hat{c}_{n-1}
    -  2M n (n-2) \hat{c}_{n-2}.
\label{rrCR}
\end{eqnarray}
\end{widetext}
For $n<0$, $\hat{a}_n$, $\hat{b}_n$, and $\hat{c}_n$ are zero.
Eqs.~(\ref{rrAR})--(\ref{rrCR}) allow $\hat{a}_0$, $\hat{b}_0$, and
$\hat{c}_0$ to be chosen freely. Other early coefficients in the
sequences are given by
\begin{eqnarray}
\pm i\Omega \hat{a}_1 &=& -(1 \pm i\Omega M)\hat{a}_0 - M \hat{b}_0/R,
\\
\pm i\Omega \hat{b}_1 &=& - M\Omega^2 R \hat{a}_0
- (2 \pm i\Omega M) b_0 + 2 c_0, \\
\pm i\Omega \hat{c}_1 &=& b_0 - c_0.
\end{eqnarray}

The set of homogeneous solutions to Eqs.~(\ref{5.24})--(\ref{5.26}),
inside and outside the orbit, forms a six dimensional linear vector
space. The six amplitudes
\[
a_0, \quad a_1, \quad c_0, \quad \hat{a}_0, \quad \hat{b}_0, \quad
\hat{c}_0
\]
determine one complete homogeneous solution and may be considered to
be the ``components'' of any member of this vector space.  The six
amplitudes that generate the particular solution which satisfies the
matching conditions of Eqs.~(\ref{5.30})--(\ref{5.33}) therefore
identify the member of the vector space that corresponds to the
desired gauge transformation.

\subsection*{Numerical integration of the ABC equations}

The numerical integration of the homogeneous versions of
Eqs.~(\ref{5.24})--(\ref{5.26}) is performed by first choosing
starting points $r_{\rm min}$ and $r_{\rm max}$ which are close
enough to their limiting values, respectively $2M$ and infinity, that
the expansions and recursion relations (\ref{5.34})--(\ref{rrCL}) and
(\ref{5.40})--(\ref{rrCR}) provide appropriate initial conditions for
$A$, $B$, and $C$ at machine accuracy with a reasonable number of
terms in the sums (no more than 30 in our case). Also, the starting
points are chosen to be sufficiently close to $R$ that the resulting
integration to $R$ takes only a few seconds of machine time.
Satisfying these two requirements simultaneously, both inside and
outside the orbit, is not difficult in practice.

The integration routine requires six input parameters, the complex
amplitudes $a_0$, $a_1$, $c_0$, $\hat{a}_0$, $\hat{b}_0$, and
$\hat{c}_0$. These must be chosen so that the six jump conditions
(\ref{5.30})--(\ref{5.33}) are enforced. We have six algebraic
equations for six unknowns. We pick a set of six linearly independent
``basis solutions'', each of which has only one of the $a_0 \ldots
\hat{c}_0$ equal to 1, all other amplitudes being zero. After
integrating the basis solutions to $R$ we collect the values of $A$,
$B$, $C$, $dA/dr$, $dB/dr$ and $dC/dr$, all evaluated at $R$, in a
matrix
\[
\mathsf{M} = \left(
\begin{array}{cccccc}
-A_{1<} & -A_{2<} & -A_{3<} & A_{1>} & A_{2>} & A_{3>} \\
-B_{1<} & -B_{2<} & -B_{3<} & B_{1>} & B_{2>} & B_{3>} \\
-C_{1<} & -C_{2<} & -C_{3<} & C_{1>} & C_{2>} & C_{3>} \\
-A'_{1<} & -A'_{2<} & -A'_{3<} & A'_{1>} & A'_{2>} & A'_{3>} \\
-B'_{1<} & -B'_{2<} & -B'_{3<} & B'_{1>} & B'_{2>} & B'_{3>} \\
-C'_{1<} & -C'_{2<} & -C'_{3<} & C'_{1>} & C'_{2>} & C'_{3>}
\end{array} \right),
\]
where we use an obvious notation; for example, $A'_{1<}$ is the
value of $dA/dr$ at $r=R$ for the first of the internal basis
solutions. The required amplitudes of the six basis solutions form the
unknown column vector $\mathbf{x}$, and the column vector $\mathbf{j}$
contains the values of the discontinuities obtained from the jump
conditions (\ref{5.30})--(\ref{5.33}). After integrating the six basis
solutions, we are left to solve the system of linear equations
\begin{equation}
\mathsf{M} \mathbf{x} = \mathbf{j}
\end{equation}
for the desired amplitudes $ \mathbf{x}$ of our basis solutions; these
then combine to give us the desired solution of
Eqs.~(\ref{5.24})--(\ref{5.26}) with appropriate boundary conditions.

In our numerical work we use double-precision arithmetic and have
adopted two different ODE integration routines from Chapter~16 of
{\it Numerical Recipes} \cite{numerical-recipes:92}, the Runge-Kutta
and the Burlish-Stoer algorithms. Each of these routines contains an
accuracy parameter.  A comparison of the numerical results over a
range of values of this parameter allows us to be certain that all
digits quoted in Table~I are significant. We tested the consistency
of the integrations versus the expansions by numerically integrating
over a wide range in $r$ where the expansions give accurate
values for $A$, $B$ and $C$. The consistency of the expansion
routines with the integration routines is strong evidence that coding
errors have been eliminated. Furthermore, we have written two
independent codes, one per author, and all results were obtained
independently before they were compared with each other. The
agreement was well within the numerical errors of each code. Our
final results for $A_<^+(R)$, $B_<^+(R)$ and $C_<^+(R)$ are listed in
Table I for selected values of $R$. Results for $A_<^-(R)$,
$B_<^-(R)$ and $C_<^-(R)$ are obtained by complex conjugation.
Results for $A_>^\pm(R)$, $B_>^\pm(R)$ and $C_>^\pm(R)$ are obtained
from the jump conditions of Eq.~(\ref{5.30}).

\begin{widetext}
\begin{table*}
\caption{Computed values for the internal functions $A_<^+(R)$,
$B_<^+(R)$ and $C_<^+(R)$. The external values are obtained by
applying the jump conditions: $\mbox{Re}[A^+_>] = \mbox{Re}[A^+_<]
+ R^2$ and the imaginary parts are identical (similar statements hold
for $B$ and $C$). The functions $A^-(R)$, $B^-(R)$, and $C^-(R)$ are
obtained by complex conjugation. All digits provided are
significant. Note that we have set $M \equiv 1$ in our computations.}
\begin{ruledtabular}
\begin{tabular}{ccccccc}
$R$ & $\mbox{Re}[A^+_<]$ & $\mbox{Im}[A^+_<]$ & $\mbox{Re}[B^+_<]$ &
                 $\mbox{Im}[B^+_<]$ $R$ & $\mbox{Re}[C^+_<]$ & $\mbox{Im}[C^+_<]$ \\
\hline
 6  & -39.427067 & -0.68518043  & -28.037347 & 3.1558616 & -26.185013 & 3.8814154 \\
 7  & -52.930571 & -0.68313011  & -38.997736 & 3.6225886 & -37.151392 & 4.3105572 \\
 8  & -68.389381 & -0.66104724  & -51.991922 & 4.0456669 & -50.152019 & 4.6962809 \\
 9  & -85.826868 & -0.63522323  & -67.000673 & 4.4373057 & -65.166204 & 5.0547070 \\
 10 & -105.25284 & -0.61006840  & -84.015966 & 4.8042522 & -82.185790 & 5.3926745 \\
 11 & -126.67197 & -0.58677222  & -103.03408 & 5.1508692 & -101.20722 & 5.7139596 \\
 12 & -150.08673 & -0.56553614  & -124.05317 & 5.4802524 & -122.22887 & 6.0210382 \\
 13 & -175.49853 & -0.54624909  & -147.07229 & 5.7947312 & -145.24995 & 6.3157134 \\
 14 & -202.90822 & -0.52871248  & -172.09095 & 6.0961288 & -170.27013 & 6.5993856 \\
 15 & -232.31636 & -0.51271828  & -199.10891 & 6.3859127 & -197.28925 & 6.8731864 \\
 16 & -263.72332 & -0.49807505  & -228.12606 & 6.6652898 & -226.30729 & 7.1380554 \\
 17 & -297.12936 & -0.48461496  & -259.14236 & 6.9352693 & -257.32427 & 7.3947870 \\
 18 & -332.53466 & -0.47219375  & -292.15781 & 7.1967063 & -290.34024 & 7.6440626 \\
 19 & -369.93935 & -0.46068823  & -327.17244 & 7.4503337 & -325.35525 & 7.8864727 \\
 20 & -409.34355 & -0.44999342  & -364.18629 & 7.6967858 & -362.36939 & 8.1225343 \\
 25 & -636.35940 & -0.40594758  & -579.24539 & 8.8386535 & -577.42896 & 9.2219041 \\
 30 & -913.37000 & -0.37286764  & -844.29131 & 9.8621836 & -842.47459 & 10.213754 \\
 35 & -1240.3777 & -0.34683600  & -1159.3279 & 10.797069 & -1157.5106 & 11.123814 \\
 40 & -1617.3835 & -0.32564845  & -1524.3578 & 11.662460 & -1522.5399 & 11.969054 \\
 45 & -2044.3881 & -0.30796022  & -1939.3826 & 12.471537 & -1937.5641 & 12.761341 \\
 50 & -2521.3918 & -0.29289834  & -2404.4036 & 13.233822 & -2402.5846 & 13.509349 \\
 55 & -3048.3949 & -0.27986801  & -2919.4216 & 13.956447 & -2917.6021 & 14.219636 \\
 60 & -3625.3975 & -0.26844793  & -3484.4373 & 14.644917 & -3482.6173 & 14.897304 \\
 65 & -4252.3997 & -0.25833015  & -4099.4510 & 15.303582 & -4097.6306 & 15.546405 \\
 70 & -4929.4016 & -0.24928356  & -4764.4632 & 15.935949 & -4762.6423 & 16.170225 \\
 75 & -5656.4033 & -0.24113082  & -5479.4740 & 16.544892 & -5477.6528 & 16.771470 \\
 80 & -6433.4048 & -0.23373328  & -6244.4838 & 17.132801 & -6242.6622 & 17.352398 \\
 85 & -7260.4061 & -0.22698069  & -7059.4926 & 17.701689 & -7057.6707 & 17.914916 \\
 90 & -8137.4073 & -0.22078417  & -7924.5006 & 18.253269 & -7922.6784 & 18.460652 \\
 95 & -9064.4083 & -0.21507114  & -8839.5079 & 18.789014 & -8837.6855 & 18.991011 \\
 100& -10041.409 & -0.20978162  & -9804.5146 & 19.310200 & -9802.6919 & 19.507212
\end{tabular}
\end{ruledtabular}
\end{table*}
\end{widetext}

\subsection*{Calculation of the self-acceleration}

Substitution of Eqs.~(\ref{5.27})--(\ref{5.29}) into
Eqs.~(\ref{5.21})--(\ref{5.23}), these into Eqs.~(\ref{5.19}),
(\ref{5.20}), and finally, these into Eq.~(\ref{5.12}) yields
\begin{eqnarray*}
h_{\alpha\beta} &=& h^{\rm s}_{\alpha\beta} - \Bigl( \bigl[
\xi_{\alpha} \bigr] r_{,\beta} + r_{,\alpha} \bigl[
  \xi_{\beta} \bigr] \Bigr) \delta(r-R)
\\ & & \mbox{}
- \Bigl( \xi^<_{\alpha;\beta} + \xi^<_{\beta;\alpha} \Bigr)
\Theta(R-r)
\\ & & \mbox{}
- \Bigl( \xi^>_{\alpha;\beta} + \xi^>_{\beta;\alpha} \Bigr)
\Theta(r-R)
\end{eqnarray*}
in an obvious notation; for example $\xi^<_\alpha$ is the internal
($r<R$) solution to Eq.~(\ref{5.13}), constructed from $A_<(r)$,
$B_<(r)$, and $C_<(r)$. The first three terms on the right-hand side
appear to be singular, but it is easy to check that by virtue of
Eqs.~(\ref{5.8})--(\ref{5.11}) and (\ref{5.30}), the factors
multiplying $\delta(r-R)$ are all zero. We therefore have
\begin{equation}
h_{\alpha\beta} = - \Bigl( \xi^<_{\alpha;\beta}
+ \xi^<_{\beta;\alpha} \Bigr) \Theta(R-r)
- \Bigl( \xi^>_{\alpha;\beta} + \xi^>_{\beta;\alpha}
\Bigr) \Theta(r-R).
\label{5.46}
\end{equation}
The jump conditions (\ref{5.31})--(\ref{5.33}) also enforce
\[
\bigl[ \xi_{\alpha;\beta} + \xi_{\beta;\alpha} \bigr] = 0,
\]
and we see that in the Lorenz gauge, the metric perturbation is
continuous at $r=R$. Equation (\ref{5.46}) also reveals that the
internal ($r<R$) and external ($r>R$) forms of $h_{\alpha\beta}$ are
obtained by a pure gauge transformation. The internal and external
transformations, however, are distinct, and the perturbation is
not globally pure gauge.

Differentiation of $h_{\alpha\beta}$ gives
\begin{eqnarray*}
h_{\alpha\beta;\gamma} &=&
- \bigl[ \xi_{\alpha;\beta} +
  \xi_{\beta;\alpha} \bigr] r_{,\gamma} \delta(r-R)
\\ & & \mbox{}
- \Bigl( \xi^<_{\alpha;\beta\gamma}
+ \xi^<_{\beta;\alpha\gamma} \Bigr) \Theta(R-r)
\\ & & \mbox{}
- \Bigl( \xi^>_{\alpha;\beta\gamma}
+ \xi^>_{\beta;\alpha\gamma} \Bigr) \Theta(r-R).
\end{eqnarray*}
Once more the singular terms vanish and we end up with the nonsingular
(but discontinuous) tensor
\begin{eqnarray}
h_{\alpha\beta;\gamma} &=&
- \Bigl( \xi^<_{\alpha;\beta\gamma}
+ \xi^<_{\beta;\alpha\gamma} \Bigr) \Theta(R-r)
\nonumber \\ & & \mbox{}
- \Bigl( \xi^>_{\alpha;\beta\gamma}
+ \xi^>_{\beta;\alpha\gamma} \Bigr) \Theta(r-R).
\label{5.47}
\end{eqnarray}
This can now be substituted into Eq.~(\ref{1.1}) to obtain the $l=1$,
even-parity contribution to the self-acceleration; the calculation
also involves the particle's velocity vector, $u^\mu =
(1-3M/R)^{-1/2}[1,0,0,\Omega]$. In the notation of Eq.~(\ref{1.3}), we
have
\begin{equation}
a[l=1; \mbox{even}] = -3 m \tilde{E} \frac{R-2M}{R^4(R-3M)}\,
\mbox{Re}\bigl[ B^+(R) \bigr],
\label{5.48}
\end{equation}
which must be evaluated on either side of $r=R$. To arrive at
Eq.~(\ref{5.48}) we have used the property that $B^-(R)$ is the
complex conjugate of $B^+(R)$, so that $B^+(R) + B^-(R) =
2\mbox{Re}[B^+(R)]$. That the acceleration vector depends only on the
radial component of $\xi_\alpha$ is a consequence of the facts that
the acceleration is pure gauge (in the sense given above) and that
the motion is circular. The curves displayed in Figs.~1 and 2 were
obtained by substituting the numerical results of Table I into
Eq.~(\ref{5.48}).

The $l=1$, even-parity contribution to the self-acceleration takes
different values depending on whether $B^+(R)$ is evaluated from
inside or outside the orbit. By virtue of Eq.~(\ref{5.30}), its jump
across the orbit is given by
\begin{equation}
\bigl[a[l=1; \mbox{even}] \bigr] = -3 m \tilde{E}
\frac{R-2M}{R^2(R-3M)}.
\label{5.49}
\end{equation}
When $R \gg M$, this agrees with the Newtonian result of
Eq.~(\ref{2.18}).

\subsection*{Self-acceleration in the post-Newtonian limit}

While we have not been able to find exact analytic solutions to
Eqs.~(\ref{5.24})--(\ref{5.26}), it is possible to make some progress
by linearizing the equations with respect to $M$. Solutions to these
equations are then post-Newtonian approximations to the exact,
numerically obtained solutions. We now set out to obtain these
approximations, and to compare them with the numerical results.

After linearization --- recall that $\Omega^2 = M/R^3$ is linear in
$M$ --- the homogeneous equations become
\begin{equation}
\frac{d^2 A}{dr^2} + \biggl( \Omega^2 - \frac{2}{r^2} - \frac{4M}{r^3}
\biggr)A - \frac{2M/R}{r^2} B = 0,
\label{5.50}
\end{equation}
\begin{equation}
\frac{d^2 B}{dr^2} + \biggl( \Omega^2 - \frac{4}{r^2} - \frac{4M}{r^3}
\biggr)B + \frac{4}{r^2} C = 0,
\label{5.51}
\end{equation}
and
\begin{equation}
\frac{d^2 C}{dr^2} + \frac{2M}{r^2} \frac{dC}{dr}
+ \biggl( \Omega^2 - \frac{2}{r^2} - \frac{4M}{r^3}
\biggr)C + \biggl( \frac{2}{r^2} + \frac{4M}{r^3} \biggr) B = 0,
\label{5.52}
\end{equation}
where we have omitted the $\pm$ labels for ease of notation. The
jump conditions reduce to $[A] = [B] = [C] = R^2$ and
\begin{eqnarray}
\biggl[ \frac{dA}{dr} \biggr] &=& 2R \biggl(1 + \frac{2M}{R} \biggr),
\nonumber \\
\biggl[ \frac{dB}{dr} \biggr] &=& R \biggl(1 + \frac{M}{R} \biggr),
\label{5.53} \\
\biggl[ \frac{dC}{dr} \biggr] &=& R \biggl(1 - \frac{2M}{R} \biggr).
\nonumber
\end{eqnarray}
We notice that the equations for $B$ and $C$ decouple from the
equation for $A$. In the sequel we will construct solutions to the $B$
and $C$ equations, and leave $A(r)$ undetermined; for the purposes of
calculating the self-acceleration, only $B(r)$ is required. Our
solutions will satisfy outgoing-wave boundary conditions at $r \to
\infty$, so that in the following, $B(r) \equiv B^+(r)$ and $C(r)
\equiv C^+(r)$.

To decouple the $B$ and $C$ equations we introduce the new dependent
variables $\psi_- = [B - (1-M/r)C]/R^2$ and $\psi_+ = [\frac{1}{3} B
+ \frac{2}{3} (1-M/r)C]/R^2$, such that
\begin{equation}
B = R^2 \biggl( \psi_+ + \frac{2}{3} \psi_- \biggr)
\label{5.54}
\end{equation}
and
\begin{equation}
C = R^2 \biggl( 1 + \frac{M}{r} \biggr)
\biggl( \psi_+ - \frac{1}{3} \psi_- \biggr).
\label{5.55}
\end{equation}
Away from $r=R$, these functions satisfy the differential equations
\begin{equation}
\psi_-'' + \biggl(1 - \frac{6}{z^2} - \frac{6v^3}{z^3} \biggr) \psi_-
= 0
\label{5.56}
\end{equation}
and
\begin{equation}
\psi_+'' + \psi_+ = 0,
\label{5.57}
\end{equation}
where we have introduced the rescaled independent variable $z=\Omega
r$ and the small quantity $v^3 = M \Omega = (M/R)^{3/2}$; a prime
indicates differentiation with respect to $z$. In terms of the new
variables, the jump conditions become
\begin{equation}
\bigl[ \psi_- \bigr] = v^2, \qquad
\bigl[ \psi_-' \bigr] = 3 v
\label{5.58}
\end{equation}
and
\begin{equation}
\bigl[ \psi_+ \bigr] = 1 - \frac{2}{3} v^2, \qquad
\bigl[ \psi_+' \bigr] = \frac{1}{v} \Bigl( 1 - v^2 \Bigr);
\label{5.59}
\end{equation}
matching is carried out at $z=v$.

To find solutions to Eq.~(\ref{5.56}) we use the fact that $v$ is
small and write
\begin{equation}
\psi_- = \psi_0 + v^3 \psi_1 + O(v^6).
\label{5.60}
\end{equation}
Substitution into Eq.~(\ref{5.56}) yields an equation for $\psi_0$,
\begin{equation}
\psi_0'' + \biggl(1 - \frac{6}{z^2} \biggr) \psi_0 = 0,
\label{5.61}
\end{equation}
and another equation for $\psi_1$,
\begin{equation}
\psi_1'' + \biggl(1 - \frac{6}{z^2} \biggr) \psi_1
= \frac{6}{z^3} \psi_0.
\label{5.62}
\end{equation}
We first solve these equations in the domain $z < v$. Among all
possible solutions to Eq.~(\ref{5.56}), we choose one which does not
diverge in the limit $z \to 0$. While this condition seems
appropriate for our purposes, it is important to understand that we
cannot fully justify it here: this choice must be introduced as an
additional assumption. The reason is as follows: Linearization of the
equations with respect to $M$ implies that Eqs.~(\ref{5.56}) and
(\ref{5.57}) apply only in the domain $r \gg M$, or $z \gg v^3$, and
this restriction prevents us from imposing a proper ingoing-wave
condition at the horizon ($r = 2M$, or $z = 2 v^3$). We must
therefore identify a suitable replacement for this boundary
condition, in the form of an asymptotic condition holding when $z$ is
restricted by $v^3 \ll z \ll v$. Previous experience
\cite{poisson-sasaki:95, leonard-poisson:97} with solving the
Regge-Wheeler equation \cite{regge-wheeler:57} in the low-frequency
limit ($M\Omega = v^3 \ll 1$) suggests that an appropriate
substitution is a regularity condition in the formal limit $z \to 0$.
This is the choice we make here, without confirmation that this
conclusion applies to the system (\ref{5.24})--(\ref{5.26}).

A regular solution to Eq.~(\ref{5.61}) is $\psi_0(z) = z j_2(z)$, or
\begin{equation}
\psi^<_0(z) = \biggl(\frac{3}{z^2} - 1 \biggr)\sin z - \frac{3}{z}
\cos z,
\label{5.63}
\end{equation}
where $j_2(z)$ is a spherical Bessel function. Substituting
Eq.~(\ref{5.63}) into Eq.~(\ref{5.62}) and integrating returns a
linear superposition of $z j_2(z)$, $z n_2(z)$, and $\psi_p(z)$, a
particular solution to the differential equation. The term involving
$z j_2(z)$ can be discarded, as it simply renormalizes the
zeroth-order solution of Eq.~(\ref{5.63}). The coefficient in front
of $z n_2(z)$ must then be chosen so as to yield a regular solution.
This gives
\begin{equation}
\psi^<_1(z) = -\frac{3}{2} \biggl( \frac{1}{z} - \frac{2}{z^3}
\biggr)\sin z + \frac{1}{2} \biggl( 1 - \frac{6}{z^2} \biggr) \cos z,
\label{5.64}
\end{equation}
and the complete interior solution to Eq.~(\ref{5.56}) is
\begin{equation}
\psi_-^<(z) = a \bigl[ \psi^<_0 + v^3 \psi^<_1 + O(v^6) \bigr].
\label{5.65}
\end{equation}
The amplitude $a$ will be determined by matching.

We next turn to the domain $z>v$ and construct an exterior solution to
Eq.~(\ref{5.56}); this will be required to satisfy an outgoing-wave
condition as $z \to \infty$. The procedure is largely the same as for
the interior solution, but is simplified by the fact that the outer
boundary is part of the domain $z \gg v^3$. An outgoing-wave
solution to Eq.~(\ref{5.61}) is $\psi_0(z) = -i z h_2^{(1)}(z)$, or
\begin{equation}
\psi^>_0(z) = \biggl( 1 + \frac{3i}{z} - \frac{3}{z^2} \biggr)
e^{iz}.
\label{5.66}
\end{equation}
Substituting this into Eq.~(\ref{5.62}) and integrating returns a
linear superposition of $z h_2^{(1)}(z)$, $z h_2^{(2)}(z)$, and
$\psi_p(z)$, a particular solution to the differential equation. As
before the term involving $z h_2^{(1)}(z)$ can be discarded, and the
term involving $z h_2^{(2)}(z)$ must also be eliminated because it
represents an incoming wave. We are left with the particular
solution,
\begin{equation}
\psi^>_1(z) = \frac{3 i}{2} \biggl( \frac{1}{z^2} + \frac{2i}{z^3}
\biggr) e^{iz}.
\label{5.67}
\end{equation}
The complete exterior solution to Eq.~(\ref{5.56}) is then
\begin{equation}
\psi_-^>(z) = b \bigl[ \psi^>_0 + v^3 \psi^>_1 + O(v^6) \bigr],
\label{5.68}
\end{equation}
and the amplitude $b$ will be determined by matching.

The constants $a$ and $b$ are determined by inserting
Eqs.~(\ref{5.65}) and (\ref{5.68}) into the jump conditions of
Eq.~(\ref{5.58}). The results are moderately complicated, and we shall
not display them here. The expressions, however, simplify once we take
into account the fact that $v$ is small. At the matching point we find
\begin{equation}
\psi_-^<(v) = -v^2 + O(v^4), \qquad
\psi_-^>(v) = O(v^4).
\label{5.69}
\end{equation}
In the interior domain ($z < v$) we can take advantage of the fact
that $z$ is formally of order $v$ to derive
\begin{equation}
\psi_-^<(z) = -(z/v)^3 v^2 + O(v^4).
\label{5.70}
\end{equation}

We now proceed with finding interior and exterior solutions to
Eq.~(\ref{5.57}). This is a much simpler task, but as we shall see,
our solutions will not be fully determined. For an interior solution
we write
\begin{equation}
\psi_+^<(z) = \alpha \bigl( \sin z - \beta v^3 \cos z \bigr),
\label{5.71}
\end{equation}
where $\alpha$ and $\beta$ are constants; the scaling of the cosine
term with $v^3$ is introduced for convenience, in anticipation of
later results. We note that this solution is regular in the formal
limit $z \to 0$, in agreement with the discussion given previously,
and that it involves two undetermined constants. For an exterior
solution we choose
\begin{equation}
\psi_+^>(z) = \gamma e^{iz},
\label{5.72}
\end{equation}
where $\gamma$ is another constant. Substitution of Eqs.~(\ref{5.71})
and (\ref{5.72}) into the jump conditions of Eq.~(\ref{5.59}) allows
us to determine $\alpha$ and $\gamma$, but $\beta$ is left over as a
free parameter. Once more the resulting expressions are too
complicated to be displayed, but they simplify for $v\ll 1$. At the
matching point we find
\begin{eqnarray}
\psi_+^<(v) &=& -1 + \biggl(\beta + \frac{2}{3} \biggr) v^2
+ O(iv^3, v^4), \nonumber \\
& & \label{5.73} \\
\psi_+^>(v) &=& \beta v^2 + O(i v^3, v^4), \nonumber
\end{eqnarray}
and in the interior domain we have
\begin{equation}
\psi_+^<(z) = -(z/v) + \biggl[ \beta + \frac{1}{2} (z/v)
+ \frac{1}{6} (z/v)^3 \biggr] v^2 + O(i v^3, v^4).
\label{5.74}
\end{equation}

Substitution of Eqs.~(\ref{5.69}) and (\ref{5.73}) into
Eqs.~(\ref{5.54}) and (\ref{5.55}) yields
\begin{equation}
B_<(R) = C_<(R) = -R^2 \biggl[ 1 - \beta \frac{M}{R}
+ O\bigl(iM\Omega,M^2/R^2\bigr) \biggr],
\label{5.75}
\end{equation}
as well as $B_>(R) = C_>(R) = \beta M R + \cdots$.
According to these results, Eq.~(\ref{5.48}) becomes
\begin{equation}
a_<[l=1; \mbox{even}] \simeq \frac{3 m \tilde{E}}{R^2}
\biggl[ 1 - (\beta-1) \frac{M}{R} \biggr]
\label{5.76}
\end{equation}
and
\begin{equation}
a_>[l=1; \mbox{even}] \simeq -\frac{3 m \tilde{E}}{R^2}
\frac{\beta M}{R}.
\label{5.77}
\end{equation}
This is compatible with the Newtonian results presented in
Eq.~(\ref{2.17}).

Substitution of Eqs.~(\ref{5.70}) and (\ref{5.74}) into
Eqs.~(\ref{5.54}) and (\ref{5.55}) gives us expressions for the
interior functions:
\begin{eqnarray}
B_<(r) &=& -R^2 \biggl\{ (r/R) - \biggl[ \beta + \frac{1}{2} (r/R)
- \frac{1}{2} (r/R)^3 \biggr] \frac{M}{R}
\nonumber \\ & & \mbox{}
+ O\bigl(iM\Omega,M^2/R^2\bigr) \biggr\}
\label{5.78}
\end{eqnarray}
and
\begin{eqnarray}
C_<(r) &=& -R^2 \biggl\{ (r/R) - \biggl[ \beta - 1 + \frac{1}{2} (r/R)
\nonumber \\ & & \hspace*{-20pt} \mbox{}
+ \frac{1}{2} (r/R)^3 \biggr] \frac{M}{R}
+ O\bigl(iM\Omega,M^2/R^2\bigr) \biggr\}.
\label{5.79}
\end{eqnarray}
These solutions are parameterized by $\beta$, which cannot be
determined here because of our lack of control over the behavior of
the solutions near $r = 2M$.

We can, however, estimate the value of $\beta$ by fitting
Eq.~(\ref{5.75}) to our numerical results. We proceed as
follows. First, we fit the expression $1 - \beta M/R + \beta'_B
(M/R)^2$ to our numerical values for $B_<(R)/(-R^2)$ in the interval
$20 \leq R/M \leq 100$; this yields $\beta = 1.9936 \pm 0.0006$ and
$\beta'_B = 4.07 \pm 0.02$. Second, we fit the expression $1 - \beta
M/R + \beta'_C (M/R)^2$ to our numerical values for $C_<(R)/(-R^2)$
restricted to the same interval; this yields $\beta = 1.9936 \pm
0.0006$ and $\beta'_C = 2.26 \pm 0.02$. Third, we fit the
expression $\beta + \beta''_B (M/R)$ to our numerical values for
$B_>(R)/(MR)$; this yields $\beta = 1.9951 \pm 0.0004$ and
$\beta''_B = -4.12 \pm 0.02$. Finally, we fit the
expression $\beta + \beta''_C (M/R)$ to our numerical values for
$C_>(R)/(MR)$; this yields $\beta = 1.9952 \pm 0.0004$ and
$\beta''_C = -2.31 \pm 0.02$. We notice an excellent consistency
among our estimates of $\beta$, and we conclude that according to
our numerical results,
\[
\beta = 1.994 \pm 0.001.
\]
It is probable that the actual value is $\beta = 2$, and that the
slight discrepancy results from a failure to include additional terms
in the expansions in powers of $M/R$. The two-parameter fits presented
here were obtained with a nonlinear least-squares Marquardt-Levenberg
algorithm, as implemented in the software {\tt gnuplot}.

The quality of the fits can be judged by comparing the
numerically-obtained functions $B_<(r)$ and $C_<(r)$ with the
post-Newtonian approximations of Eqs.~(\ref{5.78}) and
(\ref{5.79}), in which we substitute $\beta = 2$. We present this
comparison in Fig.~3 for $R=25M$. We see that the analytic expressions
are very accurate for all values of $r < R$ except near $r=2M$.

\begin{figure}
\includegraphics[angle=-90,scale=0.33]{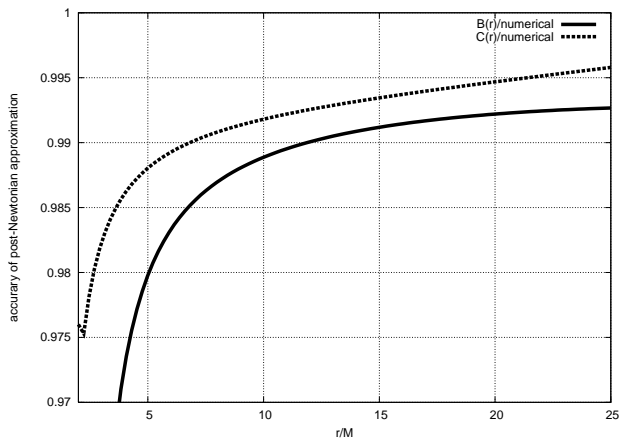}
\caption{Accuracy of the post-Newtonian expressions for the functions
$B_<(r)$ and $C_<(r)$, for $R=25M$. The solid curve is a plot of
$\mbox{Re}[B_<(r)]$, as given by Eq.~(\ref{5.78}), divided by the
numerical results listed in Table I. The dashed curve is a plot of
$\mbox{Re}[C_<(r)]$, as given by Eq.~(\ref{5.79}), divided by the
numerical results listed in Table I. In both cases we have set $\beta
= 2$. The error is estimated to be of order $(M/R)^2 \simeq 0.002$.
The plots reveal that this estimate is accurate for all values of $r$
except near $r=2M$.}
\end{figure}

\section{Discussion}

Using the tensor harmonic decomposition of Regge and Wheeler
\cite{regge-wheeler:57}, Zerilli \cite{zerilli:70} and many others
have studied the metric perturbations resulting from the geodesic
motion of a small mass in the background geometry of a Schwarzschild
black hole. Most of the attention was devoted to the radiating
modes, those with $l\ge2$, and their analysis typically involves
numerical work.

Much less attention has been garnered by the non-radiating, $l=0$ and
$l=1$ modes. In fact, the vacuum $l=1$ even-parity metric
perturbations were shown to be \textit{just gauge} by Zerilli. This
mode contains no gravitational radiation, and is usually ignored in
analyses involving gravitational perturbations of black
holes. Nonetheless, this very mode plays an important role in
self-force calculations: only the dipole mode has a Newtonian-order
contribution to the self force, as we have shown in Sec.~II. The
dipole metric perturbations cannot be ignored.

Zerilli found analytic expressions for the $l=0$ and $l=1$ metric
perturbations in a convenient gauge. The Lorenz gauge, however,
with its hyperbolic wave operator, is preferred for self-force
calculations. We, as well as Nakano, Sago, and Sasaki
\cite{nakano-etal:03}, have found analytic expressions for the
$l=0$ and odd-parity $l=1$ cases.  Our analysis of the even-parity
$l=1$ case is mostly numerical, but our procedure is robust and easy
to implement.

While the Lorenz-gauge treatment of these non-radiating modes is now
in hand, this analysis is but a small part of a complete computation
of the regularized self-acceleration, a program that was outlined in
Sec.~I. And the ultimate goal of incorporating the equations of
motion, with their corrections of order $m/M$, into a wave-generation
formalism to obtain accurate gravitational-wave templates, remains
elusive.

For example, the conservative forces discussed in this paper affect
the trajectory of the small mass at order $m/M$. But the description
of this effect inherently depends upon the choice of gauge. While the
actual observation of a gravitational-wave signal at a large distance
from the system is a gauge-independent measurement, the details of the
conversion from the self-force, as measured in the Lorentz gauge, to
the $m/M$ corrections to the wave forms, which are gauge invariant, 
are not yet known.

\begin{acknowledgments}
This work originated during the fifth Capra meeting held in May of
2002 at the Center for Gravitational Wave Physics of the Pennsylvania
State University, which is funded by the National Science Foundation
under Cooperative Agreement PHY 0114375. We thank the organizers
(Warren Anderson, Patrick Brady, Eanna Flanagan, and Lee Samuel Finn)
for their hospitality during this meeting, and the participants for
numerous discussions. The work was pursued during the sixth Capra
meeting held in June of 2003 at the Yukawa Institute for Theoretical
Physics in Kyoto, Japan, with additional support from Monbukagaku-sho
Grant-in-Aid for Scientific Research Nos. 14047212 and 14047214. We
thank the organizers (Norichika Sago, Misao Sasaki, Hideyuki Tagoshi,
Takahiro Tanaka, Hiroyuki Nakano, and Takashi Nakamura) for their
hospitality during this meeting, and the participants for additional
discussions. In addition we would like to thank Amos Ori for a very
helpful correspondence. This work was supported by the Natural
Sciences and Engineering Research Council of Canada and the National
Science Foundation under grant PHY-0245024.
\end{acknowledgments}

\appendix
\section{Monopole gauge transformation within the Lorenz gauge}

In section III we calculated the self-acceleration from the monopole metric
perturbation in the Lorenz gauge and claimed that this result was unique.
Nevertheless, our results differ from the post-Newtonian results of Nakano,
Sago, and Sasaki (NSS) \cite{nakano-etal:03}, who also work in the Lorenz
gauge. In this Appendix we outline a possible cause for the discrepancy.

Both groups begin with differing solutions in the Zerilli
gauge and then find differing gauge transformations to the Lorenz
gauge, with resulting metric perturbations that yield differing
accelerations. We ask: do our results differ from NSS by a gauge
transformation from one Lorenz gauge (ours) to another Lorenz gauge
(theirs)? An affirmative answer would invalidate our statement that
our choice of Lorenz gauge is unique. We shall instead argue that
while our results are indeed related by a gauge transformation, this
transformation takes our Lorenz gauge into another Lorenz gauge that
fails to be regular on the event horizon. We are therefore correct in
stating that our gauge choice is unique, because the gauge employed by
NSS, while appropriate for a post-Newtonian treatment, does not have a
proper relativistic generalization.

Before we investigate this matter, we note that our expression for the
metric perturbation in the Zerilli gauge has
\begin{equation}
h^{\rm Z}_{tt} = \frac{2 m \tilde{E}}{r} \Bigl( 1 - \frac{r-2M}{R-2M}
  \Bigr) \Theta(r-R),
\end{equation}
which is zero inside the orbit; this property simplifies our implementation
of the boundary conditions at the horizon. Nakano, Sago, and Sasaki, their
Eq.~(E5), have instead
\begin{equation}
h^{\rm NSS}_{tt} = \frac{2 m \tilde{E}}{r} \Bigl[
 \frac{r-2M}{R-2M}\Theta(R-r) + \Theta(r-R)
  \Bigr] ,
\end{equation}
which inside the orbit is very similar to the Newtonian result of
Eq.~(\ref{2.12}). The difference
\begin{equation}
  h^{\rm NSS}_{tt} - h^{\rm Z}_{tt} = \frac{2 m \tilde{E}}{r} \frac{r-2M}{R-2M}
\end{equation}
is a gauge transformation generated by $\xi^\alpha = [\alpha t,0,0,0]$ with
$\alpha=m\tilde{E}/(R-2M)$, so that
\begin{equation}
  -2\xi_{t;t} = h^{\rm NSS}_{tt} - h^{\rm Z}_{tt} = \frac{2 m \tilde{E}}{r}
  \frac{r-2M}{R-2M}.
\end{equation}
All other components of $\xi_{(\alpha;\beta)}$ are zero.

Now we seek an answer to our earlier questions. We assume that the
$l=0$ metric perturbation has already been obtained in one Lorenz
gauge (ours, the results appearing in Sec.~III), and we ask whether it
is possible to take it to another Lorenz gauge (which we imagine to be
a relativistic generalization of the choice made by NSS). We consider
the most general $l=0$ gauge vector
\begin{equation}
j^\alpha = [\alpha t,j(r),0,0]
\label{jdef}
\end{equation}
that keeps the metric perturbation static, where $\alpha$ is now an
arbitrary constant. The gauge transformation generated by this vector
produces a shift in the metric perturbation given by
$\Delta h_{\alpha\beta} = -2j_{(\alpha;\beta)}$. For the shifted
perturbation to satisfy the Lorenz gauge condition, the gauge vector
must satisfy the wave equation $\Box j^\alpha = 0$. This gives
\begin{equation}
j'' + \frac{2}{r} j' - \frac{2}{r^2} j = \frac{2\alpha M}{r^2 f},
\end{equation}
where $f=1-2M/r$. The general solution to this wave equation is
\begin{eqnarray}
j &=& c_1 r + c_2 \frac{M^3}{r^2} + \frac{\alpha}{9r^2}
\biggl[ 3 r^3 \ln(1-2M/r) - 3Mr^2
\nonumber\\ &&
- 12M^2 r + 44M^3 - 24M^3\ln(r/2M - 1) \biggr],
\label{gensoln}
\end{eqnarray}
where $c_1$ and $c_2$ are arbitrary, dimensionless constants. Notice
that the complicated function within the square brackets is closely
related to the function $\Gamma(r)$ defined in Eq.~(\ref{3.10}),
namely $[\ ] = -9M r^2 f \Gamma(r)$. The resulting non-zero shifts in
the metric perturbation are
\begin{eqnarray}
 \Delta h_{tt} &=& 2 \alpha f + \frac{2M}{r^2} j(r),
\nonumber\\
 \Delta h_{rr} &=& -\frac{2}{f} \frac{dj(r)}{dr}
 + \frac{2 M}{r^2 f^2} j(r),
\nonumber\\
 \Delta h_{\theta\theta} &=& \sin^2\!\theta \, \Delta h_{\phi\phi}
 = - 2r j(r).
\label{SchwPert}
\end{eqnarray}
We must now find constants $c_1$, $c_2$, and $\alpha$ which yield a
regular metric perturbation everywhere, including on the event horizon
and at infinity.

Behavior at the event horizon is easily examined in
Eddington-Finkelstein coordinates
\begin{equation}
{\cal V} = t+r+2M\ln(r/2M-1), \qquad
{\cal R} = r.
\end{equation}
The coordinates ${\cal V}$ and ${\cal R}$ are well defined everywhere
in the vicinity of the future horizon, which is located at
${\cal R} =2M$. One component of $j^\alpha$ in Eddington-Finkelstein
coordinates is
\begin{eqnarray}
  j^{\cal V} &=& j^t + j^r/f
\nonumber\\
  &=& \alpha\left[{\cal V} -{\cal R} - 2M\ln({\cal R}/2M -1)\right]
 + j({\cal R}).
\end{eqnarray}
With a substitution from Eq.~(\ref{gensoln}) it is seen that
regularity of $j^{\cal V}$ at the future horizon requires that
\begin{equation}
 2 c_1 + \frac{1}{4} c_2 + \frac{2}{9} \alpha = 0,
\label{const}
\end{equation}
and with this same condition $j^{\cal R} = O(f)$. We still need to
check the regularity of $\Delta h_{\alpha\beta}$ in
Eddington-Finkelstein coordinates,
\begin{eqnarray}
\Delta h_{\cal V V} &=& \Delta h_{tt},
 \nonumber\\
\Delta h_{\cal V R} &=& \Delta h_{tt}/f,
 \nonumber\\
\Delta h_{\cal R R} &=& \Delta h_{tt}/f^2 + \Delta h_{rr} .
\end{eqnarray}
With the choice of constants in Eq.~(\ref{const}), we see that
\begin{equation}
  \Delta h_{tt} = \alpha f [ 1 + \ln(r/2M - 1) ] + O(f^2),
\end{equation}
and
\begin{equation}
 \Delta h_{tt}/f^2 + \Delta h_{rr} = 2 \alpha + O(f).
\end{equation}
Thus $\Delta h_{\cal V V}$ and $\Delta h_{\cal R R}$ are regular on
the future horizon, but $\Delta h_{\cal V R}$ is singular if $\alpha$
is not zero. Further analysis shows that a choice of constants which
does not satisfy Eq.~(\ref{const}) only makes the shifts more
singular. Examination of behavior on the past horizon leads to the
same conclusion. The only condition on the constants that makes
$\Delta h_{\alpha\beta}$ a regular tensor field on the horizon is
$\alpha=0$ and $c_1 = -c_2/8$; but $\Delta h_{\alpha\beta}$ is then
ill behaved as $r\to\infty$. We conclude that there is no monopole
gauge transformation that simultaneously preserves the Lorenz gauge
condition and behaves properly on the event horizon and at
infinity. This confirms that our claim was true: our choice of Lorenz
gauge is indeed unique.

Nonetheless, if we set $c_1 = 0$ and make a gauge transformation with
Eq.~(\ref{jdef}), we find that the resulting change in the self-acceleration
is completely due, at first post-Newtonian order, to $j^t=\alpha t$; the
contribution from $j^r=j(r)$ appears at higher post-Newtonian order. With the
value of $\alpha$ set to $m\tilde{E}/(R-2M)$, the first post-Newtonian
contribution to the self-acceleration is $-4m M/R^3$, and it completely
accounts for the difference between our results and those of Nakano, Sago,
and Sasaki \cite{nakano-etal:03}, who present first post-Newtonian
results in their Eq.~(E19). From the discussion provided above, we
conclude that our results are related by a gauge transformation, but
that this transformation takes our Lorenz gauge into a Lorenz gauge
that fails to be regular on the event horizon. In other words, the
Lorenz gauge employed by NSS, while appropriate for their
post-Newtonian treatment, does not have a proper relativistic
generalization. 

\bibliography{motion}
\end{document}